\documentclass{aa}  

\usepackage{graphicx}
\usepackage[varg]{txfonts}
\usepackage{url}
\usepackage{aalongtable}
\usepackage{enumerate}

\usepackage{natbib}
\bibpunct{(}{)}{;}{a}{}{,} % to follow the A&A style

\usepackage{amsmath}

\begin{document}
\title{Mining the VVV: star formation and embedded clusters
  \thanks{Appendices A, B and C are only available in electronic form via http://www.edpsciences.org}}

\author{O. Solin\inst{1,3},
        L. Haikala\inst{2,3}
        \and
        E. Ukkonen\inst{1}
        }

\institute{University of Helsinki, Department of Computer Science\\
           P.O. Box 68, FI-00014 University of Helsinki, Finland\\
\email{otto.solin@helsinki.fi}
           \and
           Finnish Centre for Astronomy with ESO\\
           University of Turku, V\"ais\"al\"antie 20, FI-21500 PIIKKI\"O, Finland
           \and
           University of Helsinki, Department of Physics, Division of Geophysics and Astronomy\\
           P.O. Box 64, FI-00014 University of Helsinki, Finland
           }

\date{}

\abstract
{}
{To locate previously unknown stellar clusters from the VISTA variables in the V\'ia L\'actea Survey (VVV) catalogue data.}
{The method, fitting a mixture model of Gaussian densities and background noise using the expectation maximization algorithm to a pre-filtered NIR survey stellar catalogue data, was developed by the authors for the UKIDSS Galactic Plane Survey (GPS).}
{The search located 88 previously unknown mainly embedded stellar cluster candidates and 39 previously unknown sites of star formation in the 562 deg$^2$ covered by VVV in the Galactic bulge and the southern disk.}
{}

\maketitle
%
%________________________________________________________________

\section{Introduction}

The ESO public survey VISTA variables in the V\'ia L\'actea (VVV; \citet{VVV_DR1}) is mapping 562 deg$^2$ in the Galactic bulge and the southern disk in the \textit{ZYJHK$_s$} filters \citep{Minniti}. The VVV survey was designed to complement among others the UKIDSS Galactic Plane Survey (GPS) \citep{GPSdoc} which is mapping $\left| b \right|<5\degr$ in Galactic latitude in the northern plane.

In \citet[hereafter Paper I]{Solin} we presented an application of Gaussian mixture modelling, optimised with the Expectation Maximization (EM) algorithm \citep{EM} to automatically locate stellar clusters in the UKIDSS GPS \citep{GPSdoc}. This study applies the same method to the VVV survey first data release (DR1). The background and motivation for this work and the data mining approach to the cluster search is described in Paper I.

The search algorithm and filtering of the catalogue artefacts have been presented in detail in Paper I. The data is described in Sect. \ref{sec:theData} and the search method and results in Sects. \ref{sec:searchMethod} and \ref{sec:results}. In Sect. \ref{sec:discussion} the results, supplementary information on the cluster candidates and selected individual cluster candidates are discussed. Conclusions are drawn in Sect. \ref{sec:conclusions}.

%__________________________________________________________________

\section{The data}\label{sec:theData}
The VVV survey began in May 2010 and is expected to run in total for about five years. The DR1 catalogues contain $2.96 \times 10^8$ stellar sources detected in at least one of the five photometric bands (\textit{ZYJHK$_s$}). There are no overall limiting magnitudes as they depend strongly on crowding in the inner Galactic regions. The VVV survey is carried out by the VIRCAM (VISTA InfraRed CAMera; \citet{WFCAM}) on the VISTA (Visible and Infrared Survey Telescope for Astronomy) at ESO Paranal observatory. The VISTA Data Flow System pipeline processing and science archive are described in \citet{Irwin} and \citet{Hambly}. We have used data from the first data release, which is described in detail in \citet{VVV_DR1}. The catalogue data is used for the automated search, and the image data for visual inspection of the cluster candidate areas given by the detection algorithm. Stars brighter than $K=10^\mathrm{m}$ from the 2MASS survey are used for locating potential false positive clusters created by these bright stars (see online Appendix \ref{appB}).

%__________________________________________________________________

\section{Search method}\label{sec:searchMethod}
The search method and algorithm are described in detail in Ch. 3 in Paper I. In the following we present a short summary and outline the differencies between these two studies.

The catalogue parameter \texttt{mergedClass} classifies as stars or galaxies every object both in the WFCAM Science Archive (WSA) \citep{Hambly} catalogue data table \texttt{gpsSource} used in Paper I and the VISTA Science Archive (VSA) \citep{Cross} catalogue data table \texttt{vvvSource} used in this study. These catalogues tend to classify objects seen superposed on variable surface brightness as "galaxies". This can be utilised in the search of stellar clusters either embedded in or near molecular/dust clouds and locations of star formation. The clusters and newly formed single stars which are associated with dust clouds create variable surface brightness which broadens the stellar point spread function or creates false, extended sources. "Galaxy" in the archive parlance is more precisely described as "non-stellar" or "extended" object.

A fraction of the catalogue sources are due to data artefacts. In Paper I a few different types of artefacts in the UKIDSS survey data were addressed. In this study only false \texttt{mergedClass} $=+1$ classifications caused by diffraction patterns around bright stars are addressed. The other types addressed in Paper I are either not numerous or not at all present in the VVV survey data. Comparison between UKIDSS and VVV survey data and artefacts in them is done in online Appendices \ref{appB} and \ref{appC}.

The catalogue data table \texttt{vvvSource} contains 113 attributes for each detected object. These parameters strongly resemble those of the UKIDSS GPS. In addition to the \texttt{mergedClass} parameter we also tested in Paper I the usefulness of other parameters in our clustering effort, but ultimately both studies make use only of this star/non-stellar classifier.

As in UKIDSS the VVV classification of sources fainter than $17^\mathrm{m}$ in \textit{H} and \textit{K$_s$} as stellar/non-stellar objects is highly unreliable. These sources were filtered out from the data as they would create strong erratic background noise.

Different from Paper I we use in addition to the \textit{K} magnitude also the \textit{H} magnitude. This is because in VVV DR1 the \textit{H} magnitude is given for many tiles where the \textit{K$_s$} magnitude is missing. Indeed many candidates were found in areas where only the \textit{H} magnitude is given. In order not to loose true positives we now also do not reject any sources based on the quality error bit flags for each source detection \citep{qualityErr}. Indeed cluster candidate 12 lies in a region where the parameters \textit{hppErrBits} and \textit{ksppErrBits} often have a high value. In any case \textit{hppErrBits} and \textit{ksppErrBits} mostly fall below the value 17 and only a negligible part of the parameters \textit{hErrBits} and \textit{ksErrBits} are other than zero.

%***********************************************************************
\begin{figure*}
\centering
\includegraphics[width=\textwidth]{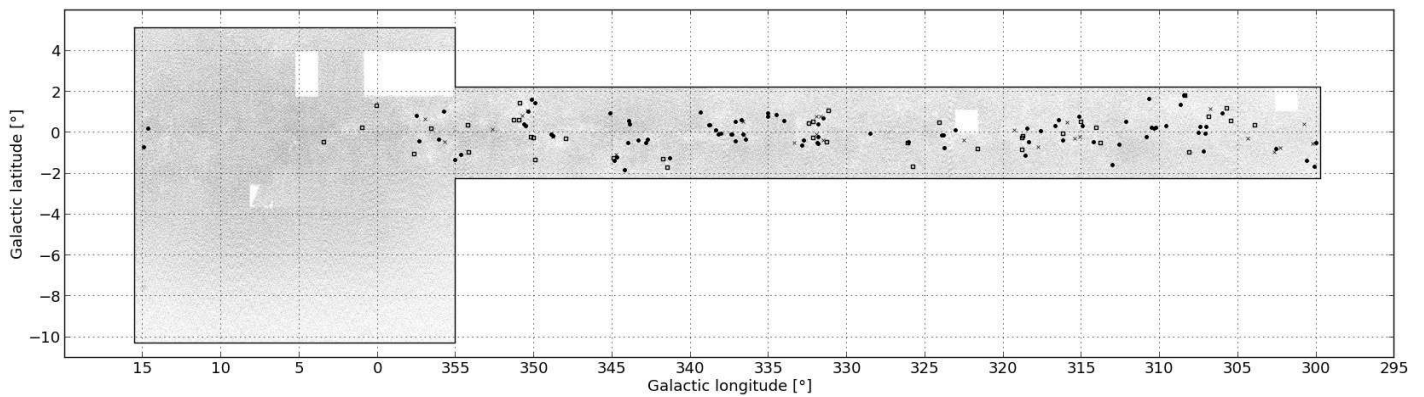}
\caption{Galactic distribution of the 88 cluster candidates (filled circles) in Table \ref{newClusters}, the 39 star formation location candidates (open circles) in Table \ref{newSFRs}, and the 26 faint nebulae (crosses) in Table \ref{newNeb}.
The grey area marks the VVV DR1 coverage in the \textit{H} or \textit{K$_s$} filter.}\label{distrIm}% 
\end{figure*}
%***********************************************************************

The VVV DR1 catalogue data table \texttt{vvvSource} contains $260,2 \times 10^6$ sources measured in the \textit{H} or \textit{K$_s$} filter with magnitude brighter than $17^\mathrm{m}$. These sources are divided according to the \texttt{mergedClass} parameter so that a negligible fraction are probable galaxies or noise, 6\% probable stars, 64\% stars, and 30\% galaxies. We end up using for the detection algorithm sources with \textit{H} or \textit{K$_s$} magnitude brighter than $17^\mathrm{m}$ and \texttt{mergedClass} $=+1$. These amount to $76,8 \times 10^6$ sources ($\sim19\%$ out of all sources  measured in the \textit{H} or \textit{K$_s$} filter in VVV DR1). Besides step \ref{magSteps} below the magnitudes listed in the VVV catalogue are in no way used in the automated search.
\\\\
The automated search proceeds in the following steps which are similar to those in Paper I where the model and its parameters are presented in detail. When \textit{K$_s$} band data is not available \textit{H} band data is used in the search.
   \begin{enumerate}
      \item The pre-filtered catalogue data is divided into smaller overlapping spatial bins of size 4\arcmin\ by 4\arcmin.
       Apart from bins at the dataset edges each bin overlaps one half of its neighbouring bins.
       4\arcmin\ by 4\arcmin\ was chosen as a suitable size for the bin based on experiments with the cluster candidates in the UKIDSS GPS list by Lucas.

      \item Remove false \texttt{mergedClass} $=+1$ classifications around bright stars as explained in Appendix A1 in Paper I.

      \item \label{magSteps} In order to track clusters with bright members the detection algorithm is run five times:
       once with all (filtered) input data and then using 80, 60, 40 and 20\% of these sources arranged in descending order of the \textit{H} or \textit{K$_s$} magnitude.

      \item The spatial coordinates are rescaled to the interval [0,1] to make all bins equally important but still allowing them to have differing means and variances.
       This step is relevant only for bins at the dataset edges and which are smaller than 4\arcmin\ by 4\arcmin.

      \item\label{initCl} In order to initialise the model parameters the data bin is divided into 16 subgrids to find the area with the highest spatial density.
       The initial value of the cluster mean $\mu$ is the center point of the subgrid with the highest density.
       The covariance matrix of the data points assigned to the subgrid with the highest density give the initial values for the cluster covariance $\Sigma$.
       The weights $\tau$ have as initial values the same value: $\tau_0 = \tau_1 = 0.5$.

      \item Each data bin is represented by a mixture model of a background component and one Gaussian cluster component as explained in Paper I.

      \item The EM-algorithm returns for each data bin a candidate cluster, i.e. an ellipse with the center point at the mean $\mu$ and
      half-axes determined by the covariance $\Sigma$.

      \item Rearrange the candidates in descending order of the Bayesian information criterion (BIC, \citet{schwarz}).

      \item Merge cluster candidates closer than one arcmin to each other.

      \item Remove from the list the cluster candidates catalogued in \citet[hereafter {[}DB2000{]}]{DB2000}; \citet[{[}DBS2003{]}]{DBS2003}; \citet[{[}BDB2003{]}]{BDB2003}; \citet[{[}BDS2003{]}]{BDS2003} (200 covered by VVV), \citet[{[}MCM2005b{]}]{MCM} (67 covered by VVV), \citet{FSR} (17 covered by VVV), Lucas (17 out of the 331 cluster candidates from UKIDSS GPS DR4 are covered also by VVV), and \citet{BorissovaVVV} (96 cluster candidates from VVV).
   \end{enumerate}

The source screening was done in exactly the manner as in Paper I (Ch. 3.3 there) with the exception that the BIC threshold was lowered to 0 (BIC can also have a negative value) in order not to loose true positives that give only a weak signal to our system. As in Paper I only a small fraction ($\sim2\%$) of the candidates given by the automated search are true cluster candidates.

We note that the center point for the cluster given by the automated search is not always exactly at the cluster center. Often during the source screening the coordinates for the candidates need to be slightly adjusted.

%__________________________________________________________________

\section{Results}\label{sec:results}

%***********************************************************************
\begin{figure*}
\centering
\includegraphics[width=\textwidth]{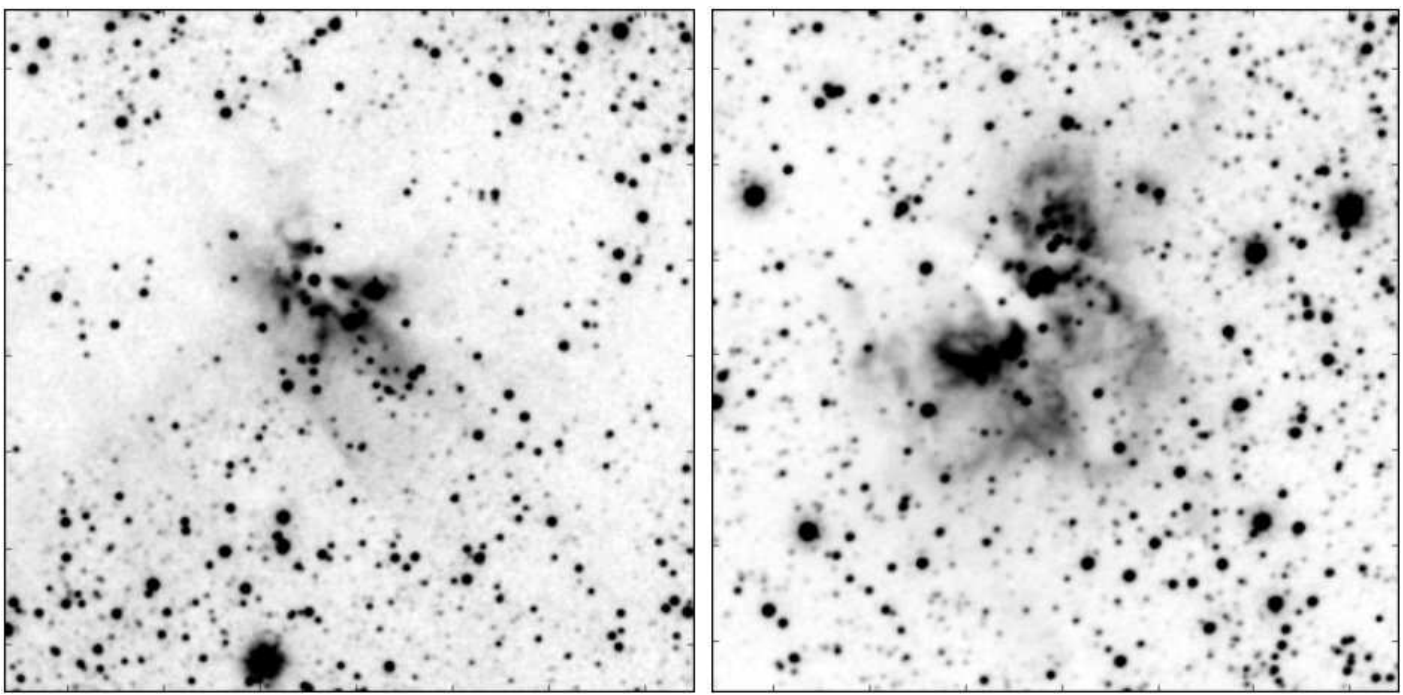}
\caption{Typical VVV \textit{K$_s$} band images of cluster candidates. On the left cluster candidate 85 and on the right cluster candidate 87. Image size is 2\arcmin\ by 2\arcmin\ and image orientation North up and East left.}\label{VVVclusters}%
\end{figure*}
%***********************************************************************

%***********************************************************************
\begin{figure*}
\centering
\includegraphics[width=\textwidth]{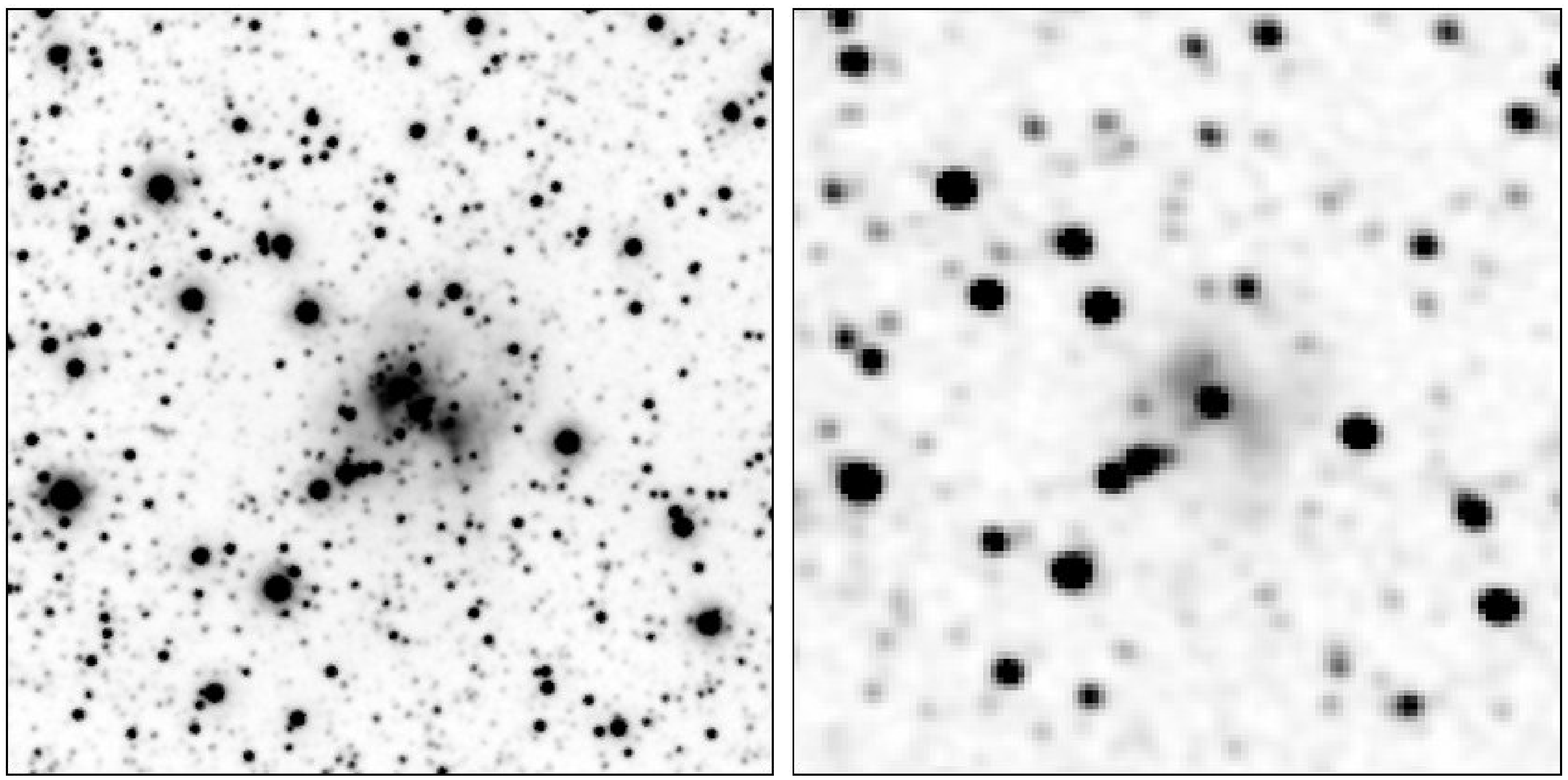}
\caption{On the left VVV \textit{K$_s$} band and on the right 2MASS image of cluster candidate 21. Image size is 2\arcmin\ by 2\arcmin\ and image orientation North up and East left.}\label{VVVand2MASS}%
\end{figure*}
%***********************************************************************

The search located 88 cluster and 39 star formation location candidates which, to our knowledge, are previously unknown. The cluster candidates are listed in Table \ref{newClusters} and the candidate locations of star formation in Table \ref{newSFRs}. In addition we list in Table \ref{newNeb} 26 faint nebulae whose nature cannot be clearly defined. The division of the candidates into these three groups is very subjective and is based on the optical appearance of the candidates and also on their SIMBAD associations:
\begin{itemize}
\item A cluster candidate has more than a few stars.
\item A location for star formation candidate has bright nebular emission but only one or a few stars. Also SIMBAD associations that are star formation indicators are considered as evidence.
\item Faint nebula candidates are similar to location for star formation candidates but are considered too weak in terms of the criteria mentioned. These sources are often in the direction of infrared dark clouds (IRDC) where a large number of other indications of star formation (IRAS, MSX and (sub)mm sources, masers and HII regions) are present. No stellar cluster can be resolved but the surface brightness (red nebulous, compact objects or only faint red surface brightness) created by putative star formation triggers the classification of the VVV \texttt{mergedClass} parameter and thus the cluster search algorithm.
\end{itemize}

The columns in the tables list (1) a running number, (2) source identification, (3,4) Galactic coordinates, (5,6) J2000.0 equatorial coordinates, (7) description of selected SIMBAD sources within 2\arcmin\ of the direction of the candidate and (8) references to selected publications in Table \ref{Pubs}.

The distribution of the candidates is shown superposed on the VVV area in Fig. \ref{distrIm}. The grey area marks the VVV DR1 coverage in the \textit{H} or \textit{K$_s$} filter. Some areas in the VVV mapping do not contain observations in neither filter. These locations are shown blank in Fig. \ref{distrIm}. The cluster candidates (Table \ref{newClusters}) are marked with filled circles, the star formation location candidates (Table \ref{newSFRs}) as open circles, and the faint nebulae (Table \ref{newNeb}) as crosses. Most candidates are in in the galactic plane outside the bulge area. This is explained by the fact that in the bulge the contamination from the field stars, despite filtering sources fainter than 17 magnitudes, is overwhelming and our method is not able to trap the clusters.

4\arcmin\ by 4\arcmin\ images in \textit{JHK$_s$} bands of the new cluster candidate areas are available in electronic form\footnote{\url{http://www.helsinki.fi/~osolin/clusters}}. Most images show clear signs of reflected light in particular in the \textit{K$_s$} band thus indicating embedded clusters or sites of star formation.

Example 2\arcmin\ by 2\arcmin\ \textit{K$_s$} band images of candidates are shown in Figs. \ref{VVVclusters} and \ref{VVVand2MASS}. Cluster candidate 85 (Fig. \ref{VVVclusters} on the left) has so far been identified as a bubble. Cluster candidate 87 (Fig. \ref{VVVclusters} on the right) has around its location an IRAS source and a millimetre source. Cluster candidate 21 (Fig. \ref{VVVand2MASS}) has so far been identified as an IRAS source and an extended 2MASS extended source (2MASX). It is also included in the list of new embedded clusters by \citet{Morales} which appeared in arXiv at the time of submission of this paper. The cluster area is shown in Fig. \ref{VVVand2MASS} on the left as a VVV image and on the right as a 2MASS image. No cluster can be seen in the 2MASS image.

Further example images of cluster candidates including their colour-colour and colour-magnitude diagrams are shown in Appendix \ref{appA}.

Besides the sources in Tables \ref{newClusters}-\ref{newNeb} the search algorithm found number of sources which are not clusters and sources which can not be clearly classified using data available presently. Some of these may be of general interest. \object{IRAS 17340-3757} and \object{IRAS 13428-6232} are two post AGB stars. Possible zone of avoidance galaxies are \object{2MASS J16353747-4459364}, \object{2MASS J16361578-4448452}, \object{2MASS J18054356-4130103} and \object{2MASS J18164114-3816136}. Object \object{2MASS J13065758-6212037} is a bright, compact source in the middle of a small dark cloud. The sources are listed in Table \ref{cat4}. Colour images of these sources are shown in online Figs. \ref{8examples_1} and \ref{8examples_2}.

%__________________________________________________________________
\section{Discussion}\label{sec:discussion}

As in Paper I SIMBAD was used to search for sources within 2\arcmin\ from the candidates in Tables \ref{newClusters}-\ref{newNeb} and \ref{cat4} with the following results (the number of sources are given in parenthesis): IRAS point source (89), MSX source (54), (sub)millimetre source (38), maser (42), outflow candidate (31) and HII region (41).

98 candidates are seen in the direction of an Infrared Dark Cloud (IRDC). IRDCs are compact, cold, dense and massive dark clouds seen in absorption against the high Galactic mid-IR surface brightness. Many IRDCs have no indication of active star formation \citep[e.g.][]{Pillai}. These clouds have been suggested to be the cold precursors to high mass star clusters \citep[e.g.][]{Rathborne}. The rest of the IRDCs are associated with typical signs of star formation, e.g. masers, IR and (sub)mm sources. In this paper some IRDCs are simply visually classified as star forming regions or faint nebulae.

Even though the search was made using only VVV stellar data the search located IRDC clouds which were not associated with clustered stars but with the faint nebulae in Table \ref{newNeb}. We argue that this surface brightness is not due to dark clouds reflecting the ambient Galactic radiation field but due to embedded star formation. The high dust extinction hides the stars and only a small fraction of the diffuse radiation produced by the star formation process is able to escape. The optical extinction of IRDCs is high even in mid-IR. Because of multiple scattering no scattered Galactic NIR radiation field is expected \citep[cf.][]{1996A&A...309..570L, 2008A&A...480..445J} and the surface brightness must have a local source. E.g. objects 21 and 23 in Table \ref{newNeb} are small angular size localised spots seen in the (\textit{H} and) \textit{K$_s$} band. Reflected light from the general ambient Galactic interstellar radiation field would rather cause a broadly distributed surface brightness at the boundaries of the densest region.

As in Paper I only IRAS point sources with fluxes typical for embedded sources in star forming clouds (a good quality flux rising from 12 microns to 60 or 100 microns) were included and cirrus-like IRAS point sources were excluded. Mostly more than one of these indicators were seen in the direction of the candidates. Only one cluster candidate (71), one star formation region candidate (23) and two objects (3 and 7) in Table \ref{cat4} had no entry within 2\arcmin\ in the SIMBAD data base.

\subsection{Notes on individual sources}\label{subsec:individ}
{\bf Cluster candidate 8} is proposed to be a young stellar object candidate \citep{AGB_YSO}. This candidate is also included in a study on embedded structures within IRDCs and other cold, massive molecular clouds \citep{Ragan}.\\
{\bf Cluster candidates 16, 17, 18 and 20} and cluster \object{[DBS2003] 131} are in the G305 star-forming complex.\\
{\bf Cluster candidates 34 and 40} are included in a study on circumstellar environments of MYSOs \citep{Wheelwright}.\\
{\bf Cluster candidate 50}: the cluster is dense but the cluster stars are not reddened.\\
{\bf Cluster candidates 59, 65 and 69} are in massive star formation complexes studied in \citet{RahmanMurray}.\\
{\bf Cluster candidate 71} has no associated SIMBAD sources within 5\arcmin. This is a dense clustering of stars. No VVV \textit{K$_s$} magnitudes are available for this field in DR1 but the colour-colour diagram using 2MASS data indicates homogeneous visual reddening of $\sim10$ magnitudes on the assumption that the cluster stars are early type stars. See online Figure \ref{ccSpecial}. \\
\onlfig{
\begin{figure*}
\includegraphics[width=\textwidth]{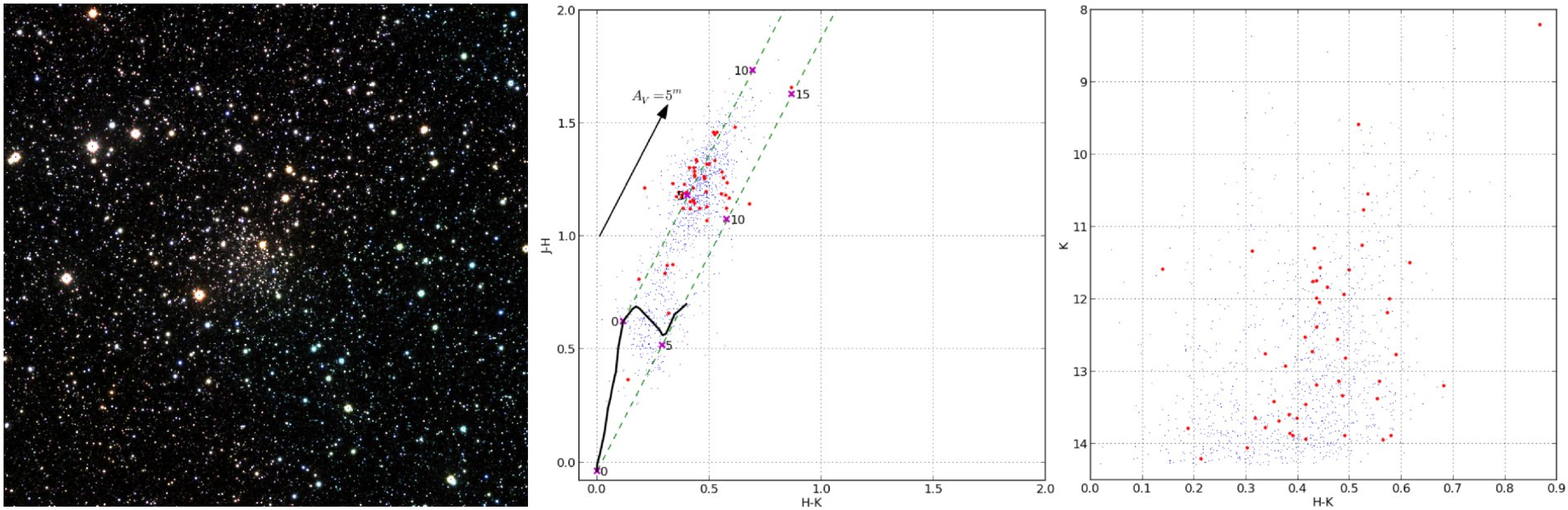}
\caption {Cluster candidate 71. On the left a false colour image produced from VVV \textit{J}, \textit{H} and \textit{K$_s$} band images. Image size is 8\arcmin\ by 8\arcmin. Image orientation is North up and East left. In the middle and on the right a colour-colour diagram and a colour-magnitude diagram drawn with 2MASS data within a 8\arcmin\ by 8\arcmin\ box around the cluster candidate. The red points are within a 0.6\arcmin\ by 0.6\arcmin\ box around the candidate.} \label{ccSpecial}
\end{figure*}
}
{\bf Cluster candidate 81} is 35\arcsec\ away from cluster \object{[MCM2005b] 88} but seems to be an individual compact cluster.\\
{\bf Faint nebula 8} is included in a study on embedded structures within IRDCs and other cold, massive molecular clouds \citep{Ragan}.\\
At least signs of {\bf cluster candidates 17, 43 and 87} can be seen also in 2MASS images, but they have not been listed as clusters in SIMBAD.\\

Many candidates are located near each other and/or near already known clusters as noted in the comments to Tables \ref{newClusters}-\ref{newNeb}.

\onlfig{
\begin{figure*}
\includegraphics[height=\textheight]{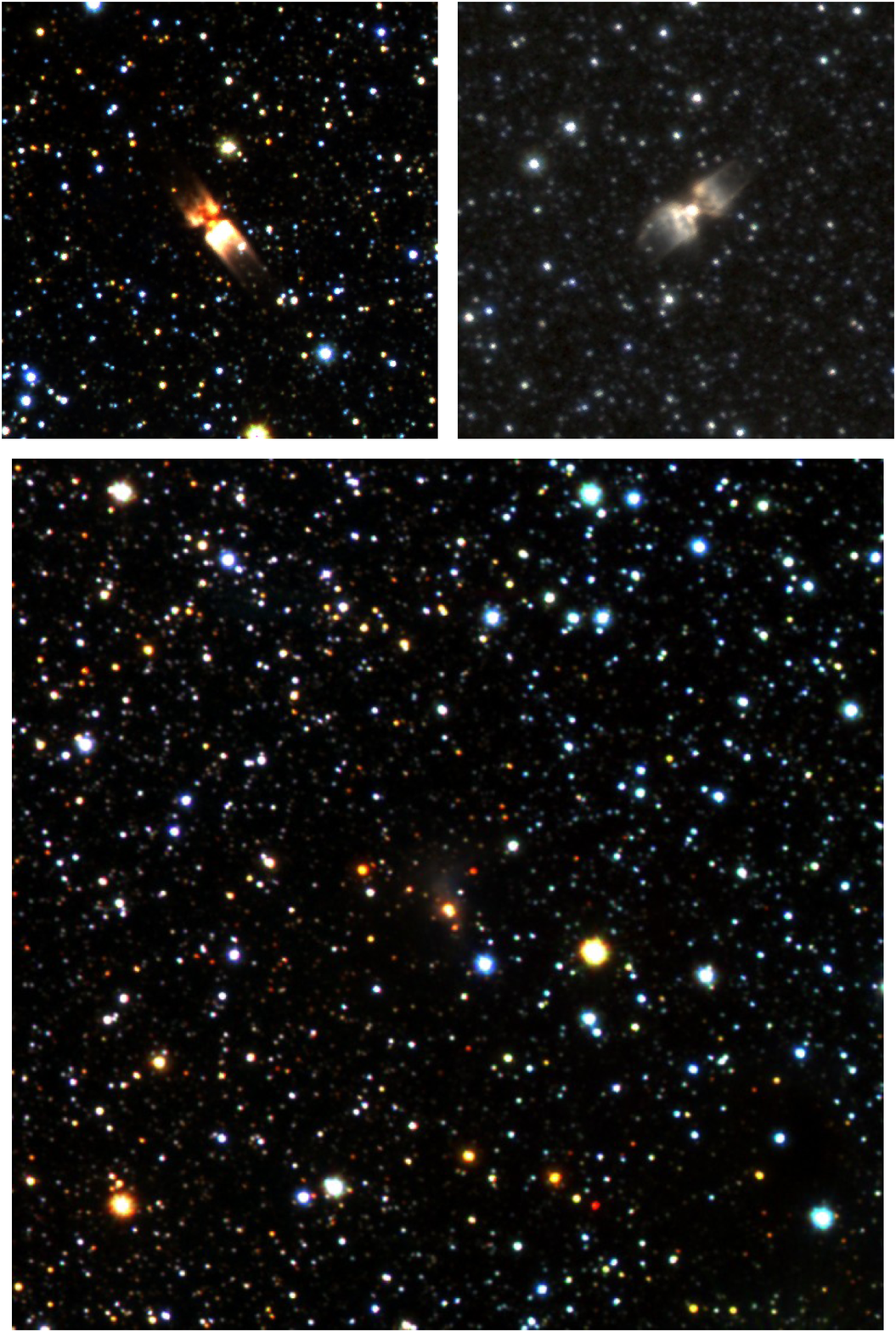}
\caption {False colour images produced from VVV \textit{J}, \textit{H} and \textit{K$_s$} band images. Image size is 2\arcmin\ by 2\arcmin\ on the upper row and 4\arcmin\ by 4\arcmin\ in the lower image. Image orientation is North up and East left. On the upper row two post AGB stars: \object{IRAS 13428-6232} and \object{IRAS 17340-3757}. Below \object{2MASS J13065758-6212037} is a bright, compact source in the middle of a small dark cloud.}\label{8examples_1}
\end{figure*}
}
\onlfig{
\begin{figure*}
\includegraphics[width=\textwidth]{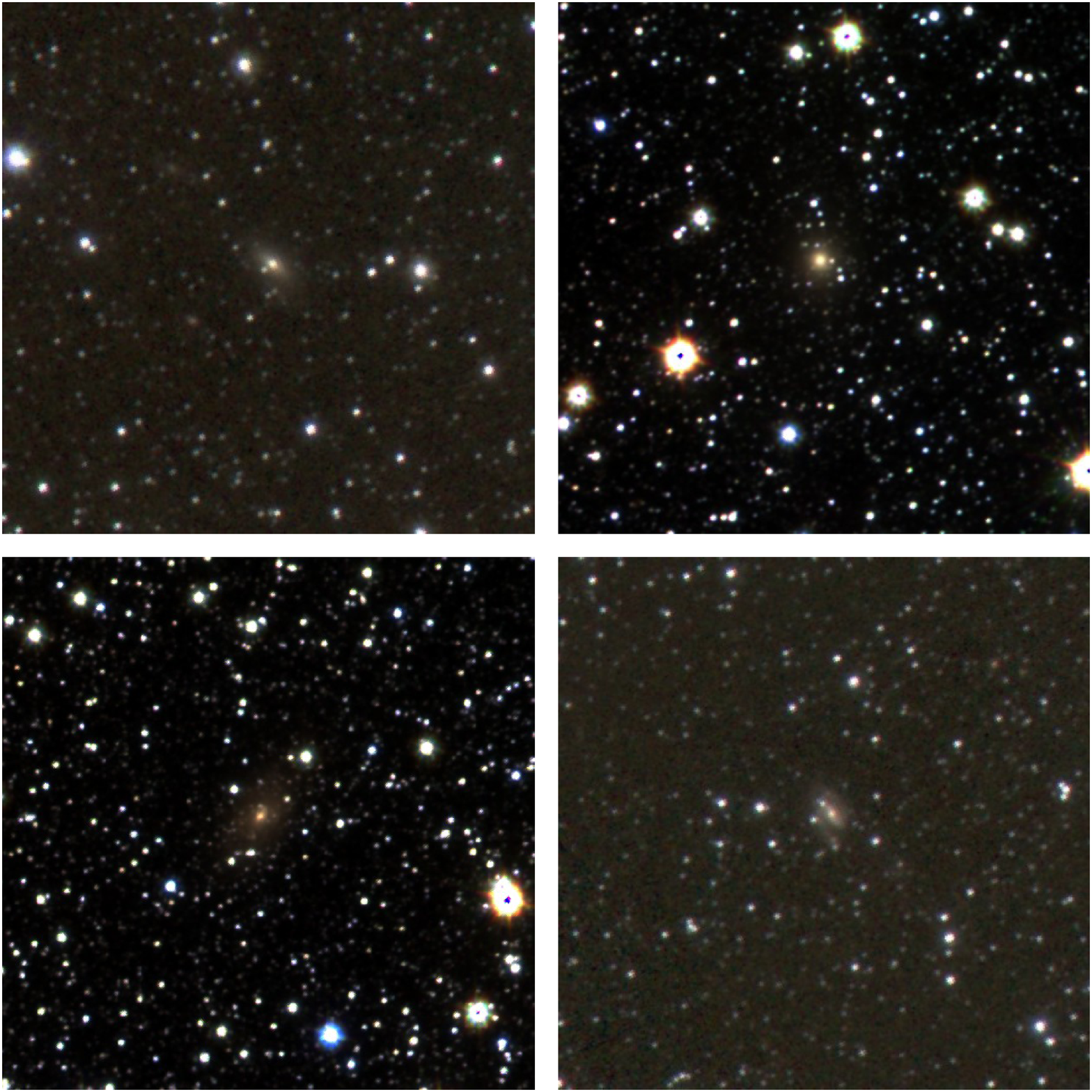}
\caption {False colour images of possible zone of avoidance galaxies produced from VVV \textit{J}, \textit{H} and \textit{K$_s$} band images. Image size is 2\arcmin\ by 2\arcmin. Image orientation is North up and East left. The sources are (clockwise from upper left corner) at the locations ($l=351.085\degr,b=-9.794\degr$), ($l=338.866\degr,b=1.669\degr$), ($l=339.075\degr,b=1.708\degr$) and ($l=354.979\degr,b=-10.189\degr$).}\label{8examples_2}
\end{figure*}
}

%______________________________________________________________

\section{Conclusions}\label{sec:conclusions}

We have applied the method developed in Paper I for the UKIDSS GPS to the VVV survey. The search from the UKIDSS GPS resulted in 137 previously unknown cluster candidates and 30 previously unknown sites of star formation. The corresponding figures for the VVV are 88 and 39 with additional 26 faint nebulae of unknown nature. There are many similarities in the results obtained from these two surveys. For both cases only a few percent of the cluster candidates produced by the automated search turn out not to be data artefacts or false positives. References 1-7 to selected publications in Table \ref{Pubs} are also listed in the equivalent table in Paper I. Particularly many of the VVV candidate SIMBAD associations are infrared dark clouds. This is not surprising as these clouds are assumed to be the forming sites of massive clusters. Similar to the UKIDDS cluster candidates in Paper I the VVV candidates do not form a homogeneous cluster sample but they vary both in size and stellar number density. Besides clustered stars (Table \ref{newClusters}) the search algorithm is triggered by locations of star formation with only one or only a few stars (Table \ref{newSFRs}) and even faint dark cloud surface brightness spots (see Table \ref{newNeb}). The number of star formation indicators seen in the direction of the candidates, and structures of surface brightness and single stars in both the UKIDSS and VVV images give confidence that most of the candidates are real entities. Like in the UKIDSS search most VVV cluster and location of star formation candidates are tightly concentrated on the Galactic plane. Most of the VVV candidates are in the disk area and only a few in the bulge where our method is not able to handle the contamination from the field stars.

\begin{acknowledgements}
This work was funded by the Finnish Ministry of Education under the project "Utilizing Finland's membership in the European Southern Observatory". This work was supported by the Academy of Finland under grants 118653 (ALGODAN) and 132291. This work uses data products from the Two Micron All Sky Survey, the United Kingdom Infrared Telescope Infrared Deep Sky Survey, and the Visible and Infrared Survey Telescope for Astronomy. This research uses the SIMBAD astronomical database service operated at CCDS, Strasbourg.
\end{acknowledgements}

\bibliographystyle{aa} % style aa.bst
\bibliography{aa22890}

\listofobjects

\tiny
\begin{longtable}{l*{8}{c}}
\caption{\label{newClusters}List of cluster candidates.}\\
\hline\hline
\# & ID & $l$ & $b$ & $\alpha$\ (J2000) & $\delta$\ (J2000) & Associated sources\tablefootmark{S} & References \\
 & & [$\degr$] & [$\degr$] & [$h$\ $m$\ $s$] & [$\degr$\ ' \ ``]  &  &  \\
\hline
\endfirsthead
\caption{continued.}\\
\hline\hline
\# & ID & $l$ & $b$ & $\alpha$\ (J2000) & $\delta$\ (J2000) & Associated sources\tablefootmark{S} & References \\
 & & [$\degr$] & [$\degr$] & [$h$\ $m$\ $s$] & [$\degr$\ ' \ ``]  &  &  \\
\hline
\endhead
\hline
\endfoot
            1 & G009.617$+$0.197 & 009.617 & $+$0.197 & 18 06 14 & $-$20 31 44 & IRAS,MSX,HII,smm,mm,Mas,IRDC,bub & 1,2,3,6,7 \\
            2 & G009.927$-$0.746 & 009.927 & $-$0.746 & 18 10 24 & $-$20 42 54 & smm,mm,IRDC,bub & 3,7 \\
            3 & G294.988$-$0.539 & 294.988 & $-$0.539 & 11 42 10 & $-$62 20 13 & \ldots & \ldots \\
            4 & G295.103$-$1.677 & 295.103 & $-$1.677 & 11 40 27 & $-$63 27 54 & IRAS & 10 \\
            5 & G295.557$-$1.378 & 295.557 & $-$1.378 & 11 45 04 & $-$63 17 46 & IRAS,MSX,2MASX & 8 \\
            6 & G297.533$-$0.823 & 297.533 & $-$0.823 & 12 03 17 & $-$63 11 20 & IRDC & 7 \\
            7 & G300.947$+$0.911 & 300.947 & $+$0.911 & 12 34 35 & $-$61 53 46 & IRAS,IRDC & 7 \\
            8 & G302.021$+$0.253 & 302.021 & $+$0.253 & 12 43 31 & $-$62 36 18 & IRAS,MSX,IRDC & 7 \\
            9 & G302.032$-$0.061 & 302.032 & $-$0.061 & 12 43 32 & $-$62 55 08 & IRAS,MSX,HII,mm,Mas,IRDC & 3,7 \\
            10 & G302.151$-$0.948 & 302.151 & $-$0.948 & 12 44 22 & $-$63 48 32 & IRAS,MSX & 8 \\
            11 & G302.390$+$0.280 & 302.390 & $+$0.280 & 12 46 44 & $-$62 35 13 & IRAS,IRDC & 7 \\
            12 & G302.487$-$0.032 & 302.487 & $-$0.032 & 12 47 32 & $-$62 54 00 & IRAS,IRDC & 7 \\
            13 & G303.405$+$1.789 & 303.405 & $+$1.789 & 12 55 21 & $-$61 04 44 & \ldots & \ldots \\
            14 & G303.641$+$1.338 & 303.641 & $+$1.338 & 12 57 23 & $-$61 31 34 & RNe & \ldots \\
            15 & G304.559$+$0.329 & 304.559 & $+$0.329 & 13 05 32 & $-$62 29 53 & IRAS,IRDC & 7 \\
            16 & G305.201$+$0.208 & 305.201 & $+$0.208 & 13 11 10 & $-$62 34 37 & MSX,HII,mm,Mas,IRDC & 3,6,7 \\
            17 & G305.350$+$0.194 & 305.350 & $+$0.194 & 13 12 28 & $-$62 34 44 & IRAS,HII,mm,Mas,IRDC,bub & 3,5,6,7 \\
            18 & G305.484$+$0.224 & 305.484 & $+$0.224 & 13 13 36 & $-$62 32 17 & MSX,IRDC & 7 \\
            19 & G305.634$+$1.648 & 305.634 & $+$1.648 & 13 13 50 & $-$61 06 22 & IRAS & \ldots \\
            20 & G305.800$-$0.250 & 305.800 & $-$0.250 & 13 16 44 & $-$62 58 52 & IRAS,MSX,HII,mm,Mas,of?,IRDC & 3,4,7 \\
            21 & G307.100$+$0.527 & 307.100 & $+$0.527 & 13 27 08 & $-$62 03 22 & IRAS,2MASX & \ldots \\
            22 & G307.560$-$0.587 & 307.560 & $-$0.587 & 13 32 31 & $-$63 05 20 & IRAS,HII,mm,IRDC & 7 \\
            23 & G307.973$-$1.595 & 307.973 & $-$1.595 & 13 37 42 & $-$64 00 47 & IRAS & \ldots \\
            24 & G309.221$-$0.463 & 309.221 & $-$0.463 & 13 46 38 & $-$62 39 32 & 2MASX,IRDC & 7 \\
            25 & G309.909$+$0.325 & 309.909 & $+$0.325 & 13 50 54 & $-$61 44 20 & of?,IRDC & 4,7 \\
            26 & G310.146$+$0.758 & 310.146 & $+$0.758 & 13 52 00 & $-$61 15 47 & IRAS,MSX,2MASX,Mas,of?,IRDC & 4,7 \\
            27 & G311.177$-$0.400 & 311.177 & $-$0.400 & 14 02 53 & $-$62 07 23 & IRAS,IRDC & 7 \\
            28 & G311.426$+$0.597 & 311.426 & $+$0.597 & 14 02 36 & $-$61 05 46 & IRAS,HII,2MASX,IRDC & 7 \\
            29 & G311.639$+$0.301 & 311.639 & $+$0.301 & 14 04 59 & $-$61 19 16 & MSX,HII,Mas,IRDC & 6,7,8 \\
            30 & G312.597$+$0.047 & 312.597 & $+$0.047 & 14 13 14 & $-$61 16 48 & IRAS,MSX,HII,mm,Mas,IRDC & 7 \\
            31 & G313.317$-$0.463 & 313.317 & $-$0.463 & 14 20 18 & $-$61 31 48 & IRAS,IRDC & 7,10 \\
            32 & G313.457$+$0.194 & 313.457 & $+$0.194 & 14 19 35 & $-$60 51 50 & MSX,HII,Mas,IRDC & 7,8 \\
            33 & G313.576$-$1.154 & 313.576 & $-$1.154 & 14 24 23 & $-$62 05 20 & IRAS & \ldots \\
            34 & G318.051$+$0.087 & 318.051 & $+$0.087 & 14 53 43 & $-$59 08 49 & IRAS,MSX,HII,Mas,of?,IRDC & 4,6,7 \\
            35 & G318.723$-$0.752 & 318.723 & $-$0.752 & 15 01 27 & $-$59 34 37 & IRAS,IRDC & 7 \\
            36 & G318.774$-$0.151 & 318.774 & $-$0.151 & 14 59 34 & $-$59 01 26 & IRAS,MSX,HII,mm,IRDC & 7 \\
            37 & G318.915$-$0.164 & 318.915 & $-$0.164 & 15 00 35 & $-$58 58 08 & HII,mm,2MASX,Mas,IRDC & 3,7 \\
            38 & G321.032$-$0.484 & 321.032 & $-$0.484 & 15 15 53 & $-$58 11 10 & MSX,HII,Mas,IRDC & 7 \\
            39 & G323.460$-$0.079 & 323.460 & $-$0.079 & 15 29 20 & $-$56 31 23 & IRAS,MSX,HII,mm,Mas,IRDC & 7,8 \\
            40 & G326.477$+$0.697 & 326.477 & $+$0.697 & 15 43 19 & $-$54 07 26 & IRAS,MSX,mm,Mas,of?,IRDC & 4,6,7,8,9 \\
            41 & G326.783$-$0.241 & 326.783 & $-$0.241 & 15 48 56 & $-$54 40 34 & IRAS,Mas,of? & 4,6 \\
            42 & G326.789$-$0.553 & 326.789 & $-$0.553 & 15 50 19 & $-$54 54 58 & IRAS,IRDC,bub & 5,7 \\
            43 & G326.794$+$0.381 & 326.794 & $+$0.381 & 15 46 21 & $-$54 10 48 & IRAS,MSX,of?,IRDC & 4,7 \\
            44 & G326.878$-$0.514 & 326.878 & $-$0.514 & 15 50 38 & $-$54 49 48 & \ldots & \ldots \\
            45 & G327.735$-$0.393 & 327.735 & $-$0.393 & 15 54 40 & $-$54 11 38 & mm,of?,IRDC & 4,7 \\
            46 & G327.809$-$0.632 & 327.809 & $-$0.632 & 15 56 07 & $-$54 19 48 & IRAS,mm,Mas,IRDC & 6,7,9 \\
            47 & G328.958$+$0.567 & 328.958 & $+$0.567 & 15 56 51 & $-$52 40 23 & IRAS,MSX,HII,IRDC & 7 \\
            48 & G329.477$+$0.842 & 329.477 & $+$0.842 & 15 58 17 & $-$52 07 41 & IRAS,MSX,IRDC & 7,8 \\
            49 & G330.020$+$0.917 & 330.020 & $+$0.917 & 16 00 38 & $-$51 43 01 & IRDC & 7 \\
            50 & G330.033$+$0.752 & 330.033 & $+$0.752 & 16 01 24 & $-$51 50 02 & \ldots & \ldots \\
            51 & G331.419$-$0.354 & 331.419 & $-$0.354 & 16 12 50 & $-$51 43 26 & HII,IRDC & 7 \\
            52 & G331.603$-$0.108 & 331.603 & $-$0.108 & 16 12 37 & $-$51 25 08 & IRAS,Mas,IRDC & 7 \\
            53 & G331.709$+$0.603 & 331.709 & $+$0.603 & 16 10 01 & $-$50 49 34 & MSX,mm,of? & 4,9 \\
            54 & G332.062$+$0.508 & 332.062 & $+$0.508 & 16 12 04 & $-$50 39 18 & IRAS,IRDC,bub & 5,7,10 \\
            55 & G332.095$-$0.421 & 332.095 & $-$0.421 & 16 16 17 & $-$51 18 22 & MSX,mm,Mas,IRDC & 7 \\
            56 & G332.294$-$0.096 & 332.294 & $-$0.096 & 16 15 46 & $-$50 56 02 & IRAS,MSX,HII,mm,Mas,of?,IRDC & 4,7 \\
            57 & G332.352$-$0.116 & 332.352 & $-$0.116 & 16 16 07 & $-$50 54 29 & Mas,of?,IRDC & 4,7 \\
            58 & G333.029$-$0.065 & 333.029 & $-$0.065 & 16 18 57 & $-$50 24 00 & MSX,2MASX,Mas,IRDC & 7 \\
            59 & G333.163$-$0.100 & 333.163 & $-$0.100 & 16 19 42 & $-$50 19 52 & MSX,HII,mm,Mas,of?,IRDC & 4,7,8 \\
            60 & G333.315$+$0.106 & 333.315 & $+$0.106 & 16 19 29 & $-$50 04 41 & MSX,Mas,of? & 4 \\
            61 & G333.725$+$0.371 & 333.725 & $+$0.371 & 16 20 08 & $-$49 36 04 & IRAS,HII,mm,IRDC & 7 \\
            62 & G333.760$+$0.364 & 333.760 & $+$0.364 & 16 20 19 & $-$49 34 55 & MSX,IRDC & 7 \\
            63 & G334.332$+$0.965 & 334.332 & $+$0.965 & 16 20 13 & $-$48 45 07 & IRAS,HII & \ldots \\
            64 & G336.290$-$1.250 & 336.290 & $-$1.250 & 16 38 10 & $-$48 51 47 & RNe & \ldots \\
            65 & G337.691$-$0.346 & 337.691 & $-$0.346 & 16 39 43 & $-$47 12 58 & IRAS,IRDC,bub & 5,7 \\
            66 & G337.784$-$0.508 & 337.784 & $-$0.508 & 16 40 48 & $-$47 15 14 & IRDC & 7 \\
            67 & G338.324$-$0.408 & 338.324 & $-$0.408 & 16 42 27 & $-$46 46 55 & of?,IRDC & 4,7 \\
            68 & G338.851$+$0.409 & 338.851 & $+$0.409 & 16 40 55 & $-$45 50 49 & IRAS,IRDC & 7 \\
            69 & G338.919$+$0.548 & 338.919 & $+$0.548 & 16 40 34 & $-$45 42 14 & MSX,Mas,of?,IRDC & 4,7,8 \\
            70 & G338.926$-$0.501 & 338.926 & $-$0.501 & 16 45 09 & $-$46 23 17 & IRAS,IRDC & 7 \\
            71 & G339.191$-$1.854 & 339.191 & $-$1.854 & 16 52 12 & $-$47 03 22 & \ldots & \ldots \\
            72 & G339.682$-$1.206 & 339.682 & $-$1.206 & 16 51 06 & $-$46 15 54 & IRAS,MSX,HII,mm,Mas & 8,9 \\
            73 & G339.801$-$1.391 & 339.801 & $-$1.391 & 16 52 22 & $-$46 17 28 & IRAS & \ldots \\
            74 & G340.071$+$0.927 & 340.071 & $+$0.927 & 16 43 16 & $-$44 35 17 & IRAS,MSX,IRDC & 7,8 \\
            75 & G343.724$-$0.183 & 343.724 & $-$0.183 & 17 00 48 & $-$42 28 26 & MSX,mm,of?,IRDC & 4,7 \\
            76 & G343.834$-$0.106 & 343.834 & $-$0.106 & 17 00 51 & $-$42 20 20 & HII,IRDC,bub & 7 \\
            77 & G343.855$-$0.099 & 343.855 & $-$0.099 & 17 00 53 & $-$42 19 08 & IRAS,HII,IRDC,bub & 7,10 \\
            78 & G344.874$+$1.435 & 344.874 & $+$1.435 & 16 57 49 & $-$40 34 08 & MSX & 8 \\
            79 & G345.119$+$1.589 & 345.119 & $+$1.589 & 16 57 59 & $-$40 16 52 & IRAS,Mas & \ldots \\
            80 & G345.327$+$1.020 & 345.327 & $+$1.020 & 17 01 00 & $-$40 28 12 & IRAS,IRDC & 7 \\
            81 & G345.489$+$0.316 & 345.489 & $+$0.316 & 17 04 28 & $-$40 46 19 & IRAS,HII,mm,Mas,IRDC & 7,9 \\
            82 & G345.590$+$0.374 & 345.590 & $+$0.374 & 17 04 33 & $-$40 39 22 & IRDC & 7 \\
            83 & G349.643$-$1.092 & 349.643 & $-$1.092 & 17 23 00 & $-$38 13 48 & IRAS,MSX,mm & 8 \\
            84 & G350.011$-$1.340 & 350.011 & $-$1.340 & 17 25 06 & $-$38 03 58 & IRAS,mm,Mas & 9 \\
            85 & G350.707$+$1.026 & 350.707 & $+$1.026 & 17 17 20 & $-$36 08 46 & bub & 9 \\
            86 & G351.043$-$0.335 & 351.043 & $-$0.335 & 17 23 50 & $-$36 38 49 & IRAS,HII,mm,bub & 9 \\
            87 & G352.314$-$0.442 & 352.314 & $-$0.442 & 17 27 48 & $-$35 39 14 & IRAS,mm & \ldots \\
            88 & G352.488$+$0.797 & 352.488 & $+$0.797 & 17 23 15 & $-$34 48 58 & IRAS,MSX,DNe & 9 \\

\end{longtable}
\tablefoot{
\tablefoottext{S}{Source classification from SIMBAD: IRDC stands for infrared dark cloud, of? for outflow candidate, bub for bubble, Mas for maser, (s)mm for (sub-)millimetre source, 2MASX for 2MASS extended source, RNe for reflection nebula and DNe for dark nebula.\\
Notes on individual sources: cluster candidate 2 is 3.1\arcmin\ away from cluster candidate 315 in the UKIDSS GPS list by Lucas, cluster candidate 16 is 3.7\arcmin\ away from [DBS2003] 131, cluster candidate 17 is 2.9\arcmin\ away from cluster \object{VVV CL022}, cluster candidate 38 and location of star formation candidate 17 are 1.8\arcmin\ apart, cluster candidate 51 is 2.1\arcmin\ away from cluster \object{VVV CL063}, cluster candidate 52 and faint nebula 18 are 2.8\arcmin\ apart, cluster candidate 56 is 4.5\arcmin\ away from cluster \object{VVV CL064}, cluster candidates 56 and 57 are 3.7\arcmin\ apart, cluster candidates 61 and 62 are 2.1\arcmin\ apart, cluster candidate 75 and faint nebula 21 are 2.4\arcmin\ apart, cluster candidates 76 and 77 are 1.3\arcmin\ apart, cluster candidate 80 and faint nebula 22 are 3.8\arcmin\ apart. Cluster candidates 7, 21, 25, 34, 36, 40, 43, 70 and 88 are included in the list of new embedded clusters by \citet{Morales} which appeared in arXiv at the time of submission of this paper.}
\tablebib{(1)~\citet{HC96}; (2) \citet{THW}; (3) \citet{HBM}; (4) \citet{EGO}; (5) \citet{CPA2006}; (6) \citet{Harju}; (7) \citet{SDC}; (8) \citet{MHL2007}; (9) \citet{CAB2011}; (10) \citet{Fontani}
}
}

\begin{table*}[h]
\caption{\label{newSFRs}List of location of star formation candidates that cannot be verified as clusters. Notes and references are as in Table \ref{newClusters}.}
\centering
         {\tiny
         \begin{tabular}{l*{8}{c}}
            \hline\hline
            \noalign{\smallskip}
\# & ID & $l$ & $b$ & $\alpha$\ (J2000) & $\delta$\ (J2000) & Associated sources\tablefootmark{S} & References \\
 & & [$\degr$] & [$\degr$] & [$h$\ $m$\ $s$] & [$\degr$\ ' \ ``]  &  &  \\
            \noalign{\smallskip}
            \hline
            \noalign{\smallskip}
            1 & G298.903$+$0.358 & 298.903 & $+$0.358 & 12 16 44 & $-$62 14 31 & of?,IRDC & 4,7 \\
            2 & G300.401$+$0.546 & 300.401 & $+$0.546 & 12 29 42 & $-$62 13 08 & IRAS,2MASX & \ldots \\
            3 & G300.720$+$1.200 & 300.720 & $+$1.200 & 12 32 50 & $-$61 35 31 & IRAS,MSX & 8 \\
            4 & G301.814$+$0.781 & 301.814 & $+$0.781 & 12 41 53 & $-$62 04 12 & IRAS & \ldots \\
            5 & G303.117$-$0.971 & 303.117 & $-$0.971 & 12 53 07 & $-$63 50 31 & IRAS,HII,2MASX & \ldots \\
            6 & G303.346$+$1.821 & 303.346 & $+$1.821 & 12 54 51 & $-$61 02 53 & IRAS & \ldots \\
            7 & G308.734$-$0.508 & 308.734 & $-$0.508 & 13 42 33 & $-$62 48 11 & IRAS & \ldots \\
            8 & G309.014$+$0.208 & 309.014 & $+$0.208 & 13 43 42 & $-$62 02 42 & IRAS & \ldots \\
            9 & G309.999$+$0.504 & 309.999 & $+$0.504 & 13 51 18 & $-$61 32 38 & of?,IRDC & 4,7 \\
            10 & G311.179$-$0.072 & 311.179 & $-$0.072 & 14 02 08 & $-$61 48 22 & MSX & 8 \\
            11 & G313.710$-$0.189 & 313.710 & $-$0.189 & 14 22 36 & $-$61 08 17 & IRAS,Mas,of? & 4 \\
            12 & G313.763$-$0.859 & 313.763 & $-$0.859 & 14 24 59 & $-$61 44 53 & IRAS,MSX,HII,Mas,of?,IRDC & 4,7 \\
            13 & G313.787$-$0.251 & 313.787 & $-$0.251 & 14 23 23 & $-$61 10 12 & MSX,IRDC & 7 \\
            14 & G316.588$-$0.809 & 316.588 & $-$0.809 & 14 46 24 & $-$60 35 46 & IRAS,MSX,IRDC & 7,8,10 \\
            15 & G319.088$+$0.460 & 319.088 & $+$0.460 & 14 59 29 & $-$58 20 13 & IRAS,2MASX,IRDC & 7 \\
            16 & G320.738$-$1.700 & 320.738 & $-$1.700 & 15 18 55 & $-$59 22 26 & DNe & \ldots \\
            17 & G321.051$-$0.507 & 321.051 & $-$0.507 & 15 16 05 & $-$58 11 46 & MSX,HII,Mas,IRDC & 6,7,9 \\
            18 & G326.145$+$1.071 & 326.145 & $+$1.071 & 15 39 58 & $-$54 01 30 & IRAS & \ldots \\
            19 & G326.270$-$0.486 & 326.270 & $-$0.486 & 15 47 11 & $-$55 11 10 & IRAS,of?,IRDC & 4,7 \\
            20 & G327.118$+$0.506 & 327.118 & $+$0.506 & 15 47 34 & $-$53 52 55 & IRAS,MSX,HII,Mas,of? & 4,6 \\
            21 & G327.131$-$0.264 & 327.131 & $-$0.264 & 15 50 54 & $-$54 28 34 & IRAS,Mas,bub & 5,10 \\
            22 & G327.403$+$0.445 & 327.403 & $+$0.445 & 15 49 19 & $-$53 45 11 & IRAS,MSX,HII,mm,Mas,of?,IRDC & 4,6,7,9 \\
            23 & G336.426$+$1.733 & 336.426 & $+$1.733 & 16 40 55 & $-$49 04 55 & \ldots & \ldots \\
            24 & G336.740$-$1.300 & 336.740 & $-$1.300 & 16 40 13 & $-$48 33 36 & \ldots & \ldots \\
            25 & G339.887$-$1.263 & 339.887 & $-$1.263 & 16 52 06 & $-$46 08 35 & IRAS,MSX,HII,Mas & 6 \\
            26 & G342.958$-$0.318 & 342.958 & $-$0.318 & 16 58 48 & $-$43 09 32 & IRAS,IRDC & 7 \\
            27 & G344.904$-$1.357 & 344.904 & $-$1.357 & 17 09 44 & $-$42 14 38 & IRAS & \ldots \\
            28 & G344.989$-$0.268 & 344.989 & $-$0.268 & 17 05 20 & $-$41 31 23 & HII,mm,IRDC & 7 \\
            29 & G345.145$-$0.217 & 345.145 & $-$0.217 & 17 05 37 & $-$41 22 05 & IRDC & 7 \\
            30 & G345.853$+$1.416 & 345.853 & $+$1.416 & 17 01 04 & $-$39 48 43 & \ldots & \ldots \\
            31 & G345.955$+$0.612 & 345.955 & $+$0.612 & 17 04 43 & $-$40 13 16 & IRAS,IRDC & 7 \\
            32 & G346.281$+$0.586 & 346.281 & $+$0.586 & 17 05 51 & $-$39 58 41 & of?,IRDC,RNe & 4,7 \\
            33 & G349.145$-$0.976 & 349.145 & $-$0.976 & 17 21 05 & $-$38 34 26 & IRAS,of?,IRDC & 4,7 \\
            34 & G349.187$+$0.344 & 349.187 & $+$0.344 & 17 15 42 & $-$37 46 48 & \ldots & \ldots \\
            35 & G351.556$+$0.205 & 351.556 & $+$0.205 & 17 23 04 & $-$35 55 08 & mm,bub & \ldots \\
            36 & G352.630$-$1.067 & 352.630 & $-$1.067 & 17 31 14 & $-$35 44 10 & IRAS,MSX,HII,mm,Mas & 9 \\
            37 & G355.031$+$1.323 & 355.031 & $+$1.323 & 17 28 00 & $-$32 24 54 & \ldots & \ldots \\
            38 & G355.937$+$0.207 & 355.937 & $+$0.207 & 17 34 46 & $-$32 16 08 & mm & \ldots \\
            39 & G358.385$-$0.484 & 358.385 & $-$0.484 & 17 43 38 & $-$30 33 58 & IRAS,MSX,HII,mm,Mas & \ldots \\

            \noalign{\smallskip}
            \hline
\end{tabular}}
\tablefoot{Location of star formation candidates 9, 12, 28 and 39 are included in the list of new embedded clusters by \citet{Morales} which appeared in arXiv at the time of submission of this paper.}
\end{table*}

\begin{table*}[h]
\caption{\label{newNeb}List of faint nebulae with unknown nature. Notes and references are as in Table \ref{newClusters}.}
\centering
         {\tiny
         \begin{tabular}{l*{8}{c}}
            \hline\hline
            \noalign{\smallskip}
\# & ID & $l$ & $b$ & $\alpha$\ (J2000) & $\delta$\ (J2000) & Associated sources\tablefootmark{S} & References \\
 & & [$\degr$] & [$\degr$] & [$h$\ $m$\ $s$] & [$\degr$\ ' \ ``]  &  &  \\
            \noalign{\smallskip}
            \hline
            \noalign{\smallskip}
            1 & G295.176$-$0.574 & 295.176 & $-$0.574 & 11 43 39 & $-$62 25 16 & HII,IRDC & 7 \\
            2 & G295.733$+$0.386 & 295.733 & $+$0.386 & 11 50 17 & $-$61 37 41 & \ldots & \ldots \\
            3 & G297.253$-$0.754 & 297.253 & $-$0.754 & 12 00 58 & $-$63 04 05 & IRAS,2MASX,IRDC & 7 \\
            4 & G297.658$-$0.975 & 297.658 & $-$0.975 & 12 04 07 & $-$63 21 40 & IRAS,HII,Mas,bub & 5 \\
            5 & G299.325$-$0.308 & 299.325 & $-$0.308 & 12 19 37 & $-$62 57 25 & IRDC,RNe & 7 \\
            6 & G301.720$+$1.121 & 301.720 & $+$1.121 & 12 41 12 & $-$61 43 37 & MSX,mm,2MASX & 8 \\
            7 & G310.076$-$0.228 & 310.076 & $-$0.228 & 13 53 23 & $-$62 14 13 & of?,IRDC & 4,7 \\
            8 & G310.375$-$0.298 & 310.375 & $-$0.298 & 13 56 01 & $-$62 13 59 & IRAS,MSX,of?,IRDC,DNe & 4,7 \\
            9 & G310.866$+$0.473 & 310.866 & $+$0.473 & 13 58 24 & $-$61 21 47 & IRDC & 7 \\
            10 & G312.742$-$0.716 & 312.742 & $-$0.716 & 14 16 26 & $-$61 57 29 & IRAS,IRDC & 7 \\
            11 & G314.274$+$0.096 & 314.274 & $+$0.096 & 14 26 08 & $-$60 40 23 & MSX,HII,Mas,IRDC & 7 \\
            12 & G317.463$-$0.398 & 317.463 & $-$0.398 & 14 51 18 & $-$59 50 42 & of?,IRDC & 4,7 \\
            13 & G326.466$-$0.381 & 326.466 & $-$0.381 & 15 47 49 & $-$54 58 55 & IRAS,MSX,HII,mm,IRDC & 7,8 \\
            14 & G326.647$+$0.749 & 326.647 & $+$0.749 & 15 44 01 & $-$53 58 44 & of?,IRDC & 4,7 \\
            15 & G326.884$-$0.104 & 326.884 & $-$0.104 & 15 48 53 & $-$54 30 22 & IRDC & 7 \\
            16 & G326.933$+$0.783 & 326.933 & $+$0.783 & 15 45 25 & $-$53 46 41 & IRAS,IRDC & 7 \\
            17 & G328.336$-$0.528 & 328.336 & $-$0.528 & 15 58 24 & $-$53 54 40 & IRDC & 7 \\
            18 & G331.561$-$0.128 & 331.561 & $-$0.128 & 16 12 30 & $-$51 27 43 & MSX,HII,mm,Mas,IRDC & 7,8 \\
            19 & G331.623$+$0.523 & 331.623 & $+$0.523 & 16 09 58 & $-$50 56 35 & IRAS,MSX,mm,of?,IRDC & 4,7,9,10 \\
            20 & G339.724$-$1.120 & 339.724 & $-$1.120 & 16 50 52 & $-$46 10 41 & IRAS,IRDC & 7 \\
            21 & G343.722$-$0.224 & 343.722 & $-$0.224 & 17 00 59 & $-$42 30 00 & IRAS,HII,IRDC & 7 \\
            22 & G345.379$+$1.056 & 345.379 & $+$1.056 & 17 01 02 & $-$40 24 25 & IRDC,bub & 5,7 \\
            23 & G345.713$+$0.815 & 345.713 & $+$0.815 & 17 03 06 & $-$40 17 24 & IRAS,MSX,mm,of?,IRDC & 4,7,9 \\
            24 & G347.617$+$0.152 & 347.617 & $+$0.152 & 17 11 48 & $-$39 09 54 & Mas,IRDC & 7 \\
            25 & G350.692$-$0.492 & 350.692 & $-$0.492 & 17 23 30 & $-$37 01 34 & IRAS,MSX,mm,Mas,IRDC,bub & 7,9 \\
            26 & G351.922$+$0.642 & 351.922 & $+$0.642 & 17 22 19 & $-$35 22 12 & IRAS,bub & \ldots \\
            \noalign{\smallskip}
            \hline
\end{tabular}}
\tablefoot{Faint nebula 24 is 3.3\arcmin\ away from cluster \object{[DBS2003] 179}. Faint nebulae 1, 5, 11, and 12 are included in the list of new embedded clusters by \citet{Morales} which appeared in arXiv at the time of submission of this paper.}
\end{table*}

\begin{table*}[ht]
\caption{\label{Pubs}List of publications referenced in Tables \ref{newClusters}, \ref{newSFRs} and \ref{newNeb}.}
\centering
         {\tiny
         \begin{tabular}{*{4}{c}}
            \hline\hline
            \noalign{\smallskip}
            \# & BibCode & Aut & Description \\
            \noalign{\smallskip}
            \hline
            \noalign{\smallskip}

            1 & 1996A\&AS..120..283H & Hofner \& Churchwell & Water maser emission of UC HII regions \\ % HC96
            2 & 2006A\&A...453.1003T & Thompson et al. & SCUBA smm survey of IRAS and UC HII regions \\ % THW2006
            3 & 2005MNRAS.363..405H & Hill et al. & mm observations of SFRs \\ % HBM2005
            4 & 2008AJ....136.2391C & Cyganowski et al. & MYSO outflow candidates \\ % EGO
            5 & 2006ApJ...649..759C & Churchwell et al. & Bubble candidates from GLIMPSE \\ % CPA2006
            6 & 1998A\&AS..132..211H & Harju et al. & SiO emission of masers \\ % HLB98
            7 & 2009A\&A...505..405P & Peretto \& Fuller & GLIMPSE IRDCs: initial conditions of stellar protocluster formation \\ % SDC
            8 & 2007A\&A...476.1019M & Mottram et al. & Mid-infrared observations of candidate massive YSOs \\ % MHL2007
            9 & 2011ApJS..195....8C & Culverhouse et al. & A Compact Source Catalog from the QUaD Galactic Plane Survey \\ % CAB2011
            10 & 2005A\&A...432..921F & Fontani et al. & Search for massive protostellar candidates \\ % Fontani

            \noalign{\smallskip}
            \hline
\end{tabular}}
\end{table*}

\onltab{
\begin{table*}[h]
\caption{\label{cat4}Sources which are not clusters but of general interest. Notes and references are as in Table \ref{newClusters}.}
\centering
         {\tiny
         \begin{tabular}{l*{8}{c}}
            \hline\hline
            \noalign{\smallskip}
\# & ID & $l$ & $b$ & $\alpha$\ (J2000) & $\delta$\ (J2000) & Associated sources\tablefootmark{S} & References \\
 & & [$\degr$] & [$\degr$] & [$h$\ $m$\ $s$] & [$\degr$\ ' \ ``]  &  &  \\
            \noalign{\smallskip}
            \hline
            \noalign{\smallskip}
            1 & G304.741$+$0.617 & 304.741 & $+$0.617 & 13 06 58 & $-$62 12 00 & IRDC & 7 \\
            2 & G309.159$-$0.593 & 309.159 & $-$0.593 & 13 46 21 & $-$62 47 56 & IRAS,MSX,2MASX & \ldots \\
            3 & G338.866$+$1.669 & 338.866 & $+$1.669 & 16 35 37 & $-$44 59 38 & \ldots & \ldots \\
            4 & G339.075$+$1.708 & 339.075 & $+$1.708 & 16 36 16 & $-$44 48 47 & \ldots & \ldots \\
            5 & G351.085$-$9.794 & 351.085 & $-$9.794 & 18 05 44 & $-$41 30 11 & \ldots & \ldots \\
            6 & G351.412$-$3.339 & 351.412 & $-$3.339 & 17 37 29 & $-$37 59 20 & \ldots & \ldots \\
            7 & G354.979$-$10.189 & 354.979 & $-$10.189 & 18 16 41 & $-$38 16 16 & \ldots & \ldots \\
            \noalign{\smallskip}
            \hline
\end{tabular}}
\end{table*}
}

\Online

\begin{appendix}

\section{Examples of cluster candidates}\label{appA}

\begin{figure*}
\centering
\includegraphics[width=\textwidth]{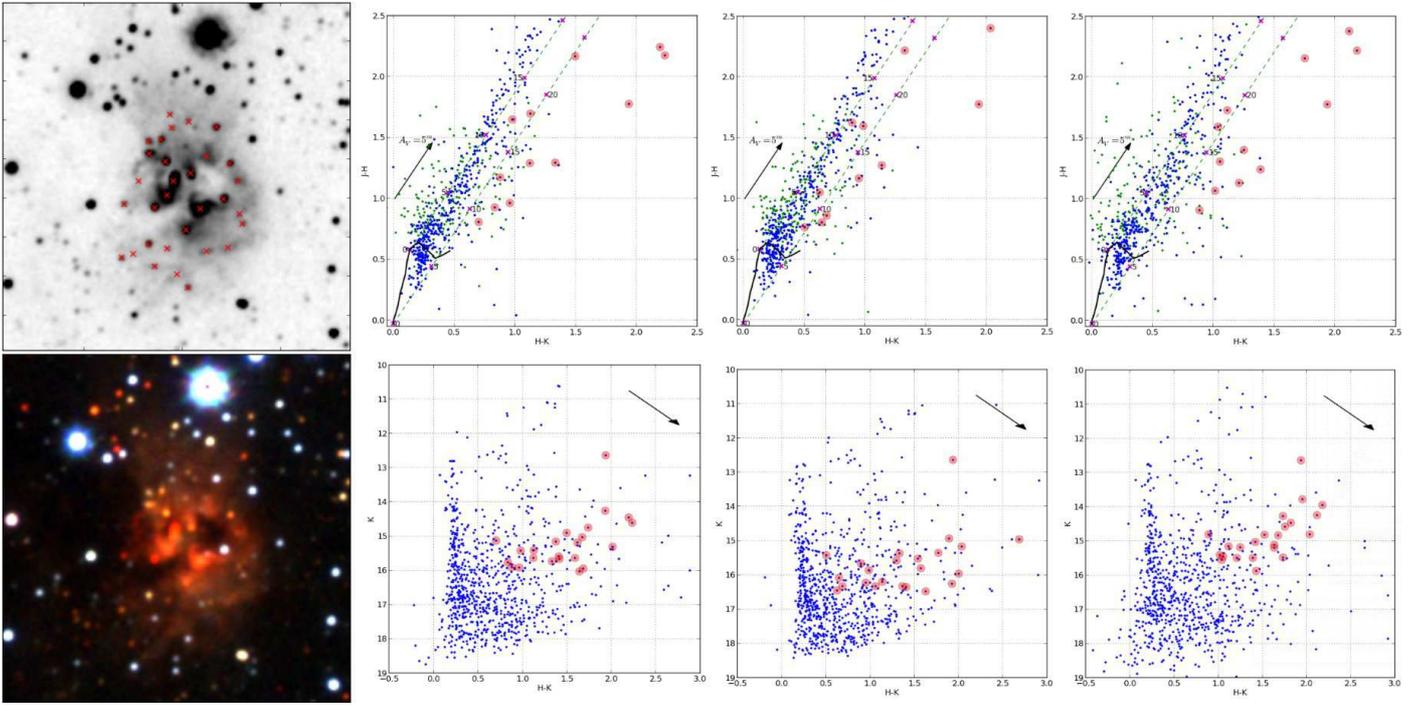}
\caption{Cluster candidate 11. The 1\arcmin\ by 1\arcmin\ false colour image of the cluster candidate is shown below the grey scale image. Image orientation is North up and East left. All the sources within a 4\arcmin\ by 4\arcmin\ box around the cluster candidate are plotted in the $(H-K, J-H)$ colour-colour and $(H-K, K)$ colour-magnitude plots. In the colour-colour plot blue dots are sources brighter than $17^\mathrm{m}$ and green dots fainter than $17^\mathrm{m}$ in \textit{K$_s$}. The colour-colour and colour-magnitude diagrams, from left to right, are plotted using the \textit{AperMag3}, \textit{AperMag1}, and the \textit{AperMag4} magnitudes, respectively. The red crosses in the 1\arcmin\ by 1\arcmin\ grey scale image and the the red filled circles in the colour plots are sources (both stellar and non-stellar) in the cluster direction brighter than $17^\mathrm{m}$ in \textit{K$_s$}.}
  \label{cc11}%
\end{figure*}

\begin{figure*}
\centering
\includegraphics[width=\textwidth]{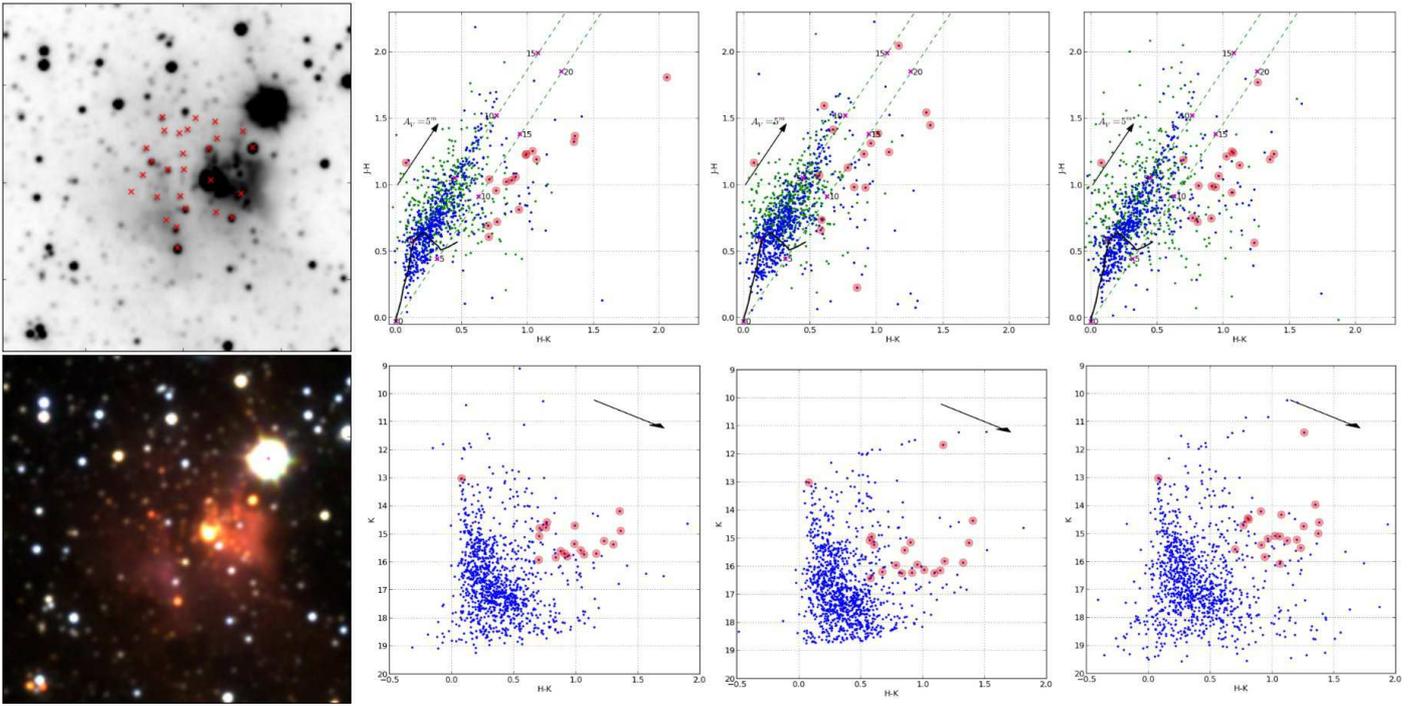}
\caption{As Fig. \ref{cc11} for cluster candidate 5.}
  \label{cc5}%
\end{figure*}

Example cluster candidates are shown in Figs. \ref{cc11} and \ref{cc5}. The different panels in the figures are as follows. On the left 1\arcmin\ by 1\arcmin\ VVV grey scale \textit{K$_s$} band and false colour image of the cluster area. The images are produced using the \textit{J}, \textit{H} and \textit{K$_s$} fits files obtained from the VSA. In the \textit{K$_s$} band image sources (both stellar and non-stellar) in the cluster direction brighter than $17^\mathrm{m}$ in \textit{K$_s$} are marked with red crosses. All the sources within a 4\arcmin\ by 4\arcmin\ box around the cluster candidate are plotted in the $(H-K, J-H)$ colour-colour and $(H-K, K)$ colour-magnitude plots. The red filled circles mark the same sources as the red crosses in the grey scale image. As in Paper I the automated search uses by default the \textit{AperMag3} magnitudes (2.0\arcsec\ aperture diameter). For the colour-colour and colour-magnitude plots we experimented also with the \textit{AperMag1} (1.0\arcsec\ aperture diameter) and \textit{AperMag4} (2.8\arcsec\ aperture diameter) extended source magnitudes. The arrow indicates an optical extinction of 5 magnitudes. In the colour-colour plot blue dots mark sources brighter than $17^\mathrm{m}$ and green dots sources fainter than $17^\mathrm{m}$ in \textit{K$_s$}. The approximate unreddened main sequence is plotted with a continuous line and approximate main sequence reddening lines are shown with dashed lines. The numbers on the reddening lines show the optical extinction in case the star originates from the early or late main sequence. The value of 1.6 for the reddening slope is the mean of all the reddening tracks in \citet{SteadHoare}.

Both example cluster candidates are in crowded fields. As in Paper I in an effort to get better precision than with AperMag3 magnitudes (2.0\arcsec\ aperture diameter) we experimented also with the AperMag1 (1.0\arcsec\ aperture diameter) and AperMag4 (2.8\arcsec\ aperture diameter) magnitudes. However the colour plots show only little variation depending on the magnitudes used.

For both cases all three colour-colour plots indicate infrared excess and all three colour-magnitude plots suggest that the cluster members are of early type.

\section{VVV and GPS catalogue artefacts}\label{appB}

Only catalogue artefacts created around bright stars coincide in the GPS and VVV surveys. This is because of the different optics of the UKIRT/WFCAM and VISTA/VIRCAM. The typical artefacts in the GPS survey are the bow-tie, beam, array edge flare, and persistence image (see App. A.1 in Paper I). Many of these artefacts either do not exist in the VVV survey or their number is greatly reduced in comparison with the GPS survey. Most importantly the automatic search algorithm used in this work does not produce as many false positive clusters due to artefacts in the VVV survey catalogue as in the UKIDSS GPS catalogue. In the latter particularly the persistence images and array edge flares produce a large number of strong false positive clusters.

\subsection{False positive clusters caused by bright stars}\label{appB1}

\begin{figure*}
\centering
\includegraphics[width=\textwidth]{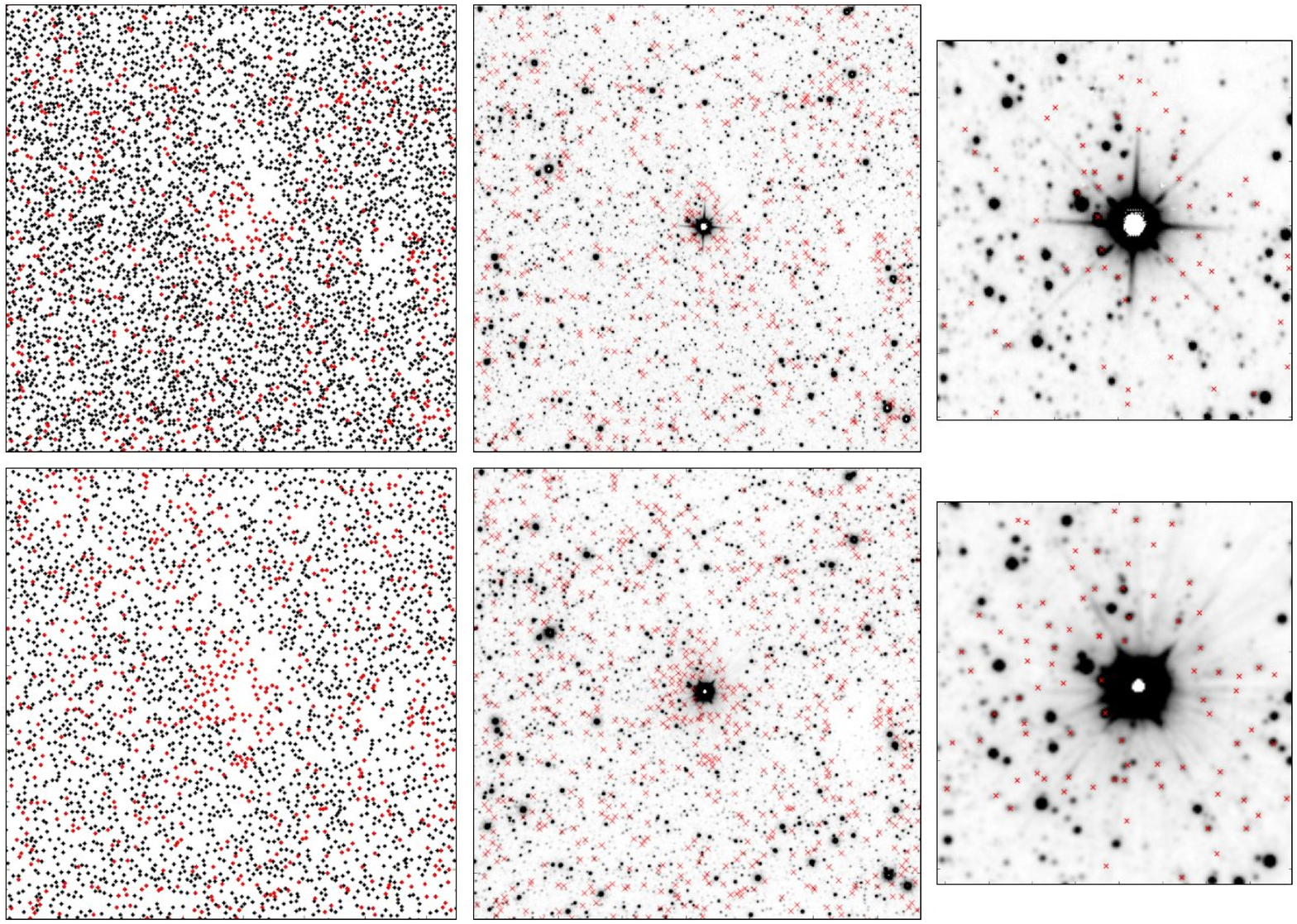}
\caption{A false positive cluster caused by the bright $K=4\fm8$ star 2MASS J17565546-2511015. On the upper row are the images and plots from UKIDSS and on the lower row from VVV. In the rightmost column are the zoomed in images of the areas around the bright central star. The red points in the catalogue plots and red crosses in the images are sources brighter than $17^\mathrm{m}$ in \textit{K} and classified as non-stellar. In the catalogue plots all other catalogue sources are plotted in black. Image orientation is North up and East left. Image size is 4\arcmin\ by 4\arcmin\ in the middle column and 1\arcmin\ by 1\arcmin\ in the rightmost column.}
  \label{brightStar}%
\end{figure*}

The most common catalogue artefact from the point of view of this work is that caused by bright stars. The extended halo around the bright stars and diffraction spikes cause varying surface brightness which either broadens the image of real stars or produces spatially extended false sources both of which are classified as non-stellar sources. An example of a false positive cluster caused by a bright star is shown in Fig. \ref{brightStar}. This field is included both in the GPS and VVV surveys. The upper three images, from left to right, in Fig. \ref{brightStar} show the catalogue plot, catalogue \textit{K$_s$} image and zoom on bright star in the \textit{K$_s$} image, respectively. The field size is 4\arcmin\ by 4\arcmin\ in the left and centre and 1\arcmin\ by 1\arcmin\ on the right. The objects classified as extended are indicated with a red cross. The corresponding VVV images are shown in the three lower images. The number of visible strong diffraction spikes is fewer in the GPS image (eight) than in the VVV image where they are weaker. The bright star produces more non-stellar classifications into the VVV catalogue than in the GPS catalogue. In the VVV catalogue all of the sources within this area are brighter than $17^\mathrm{m}$ in \textit{K} but only 85\% in the UKIDSS catalogue. Bright stars produce false positive clusters in both the GPS and VVV catalogues. For both catalogues the remedy is to discard non-stellar sources very near 2MASS stars brighter than $K=10^\mathrm{m}$ in \textit{K}. The brighter the star, the greater the distance to which it produces false classifications (see App. A.1 in Paper I).

\begin{figure*}
\centering
\includegraphics[width=\textwidth]{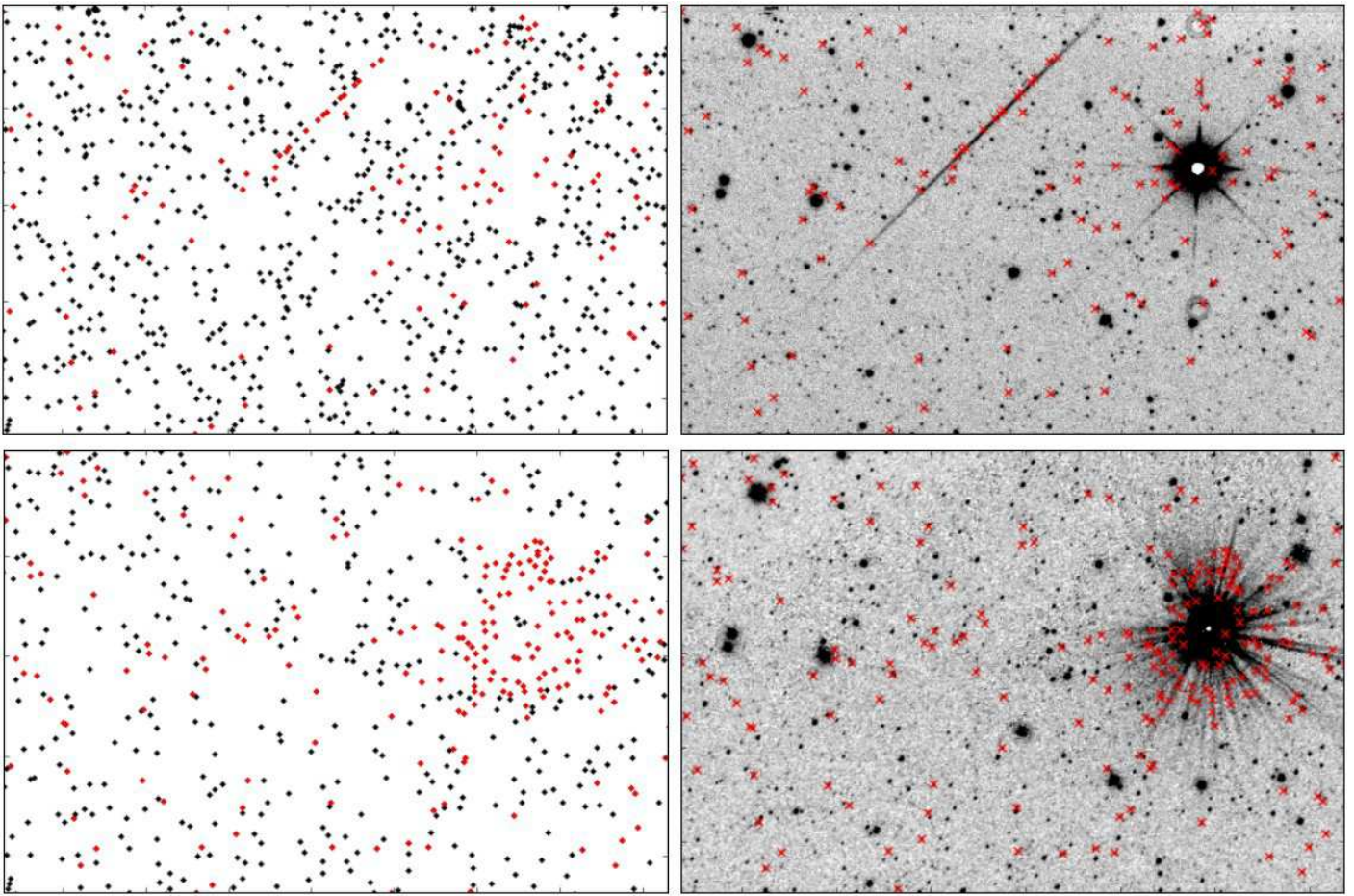}
\caption{A false positive cluster caused by the bright $K=5\fm3$ star 2MASS J18275636-3343355. On the upper row are the images and plots from UKIDSS and on the lower row from VVV. Image orientation is North up and East left. Image size is 4\arcmin\ by 2.7\arcmin. Plot markers (dots and crosses) are as in Fig. \ref{brightStar}.}
  \label{brightStar2}%
\end{figure*}

In Fig. \ref{brightStar2} the bright star produces many more non-stellar classifications into the VVV catalogue than into the UKIDSS catalogue. In the UKIDSS image there are visible two more artefacts: a beam across the image and two cross-talk images above and below the bright star. Within this area 68\% of the sources are brighter than $17^\mathrm{m}$ in \textit{K} in the VVV survey catalogue but only 48\% in the UKIDSS catalogue.

\begin{figure*}
\centering
\includegraphics[width=\textwidth]{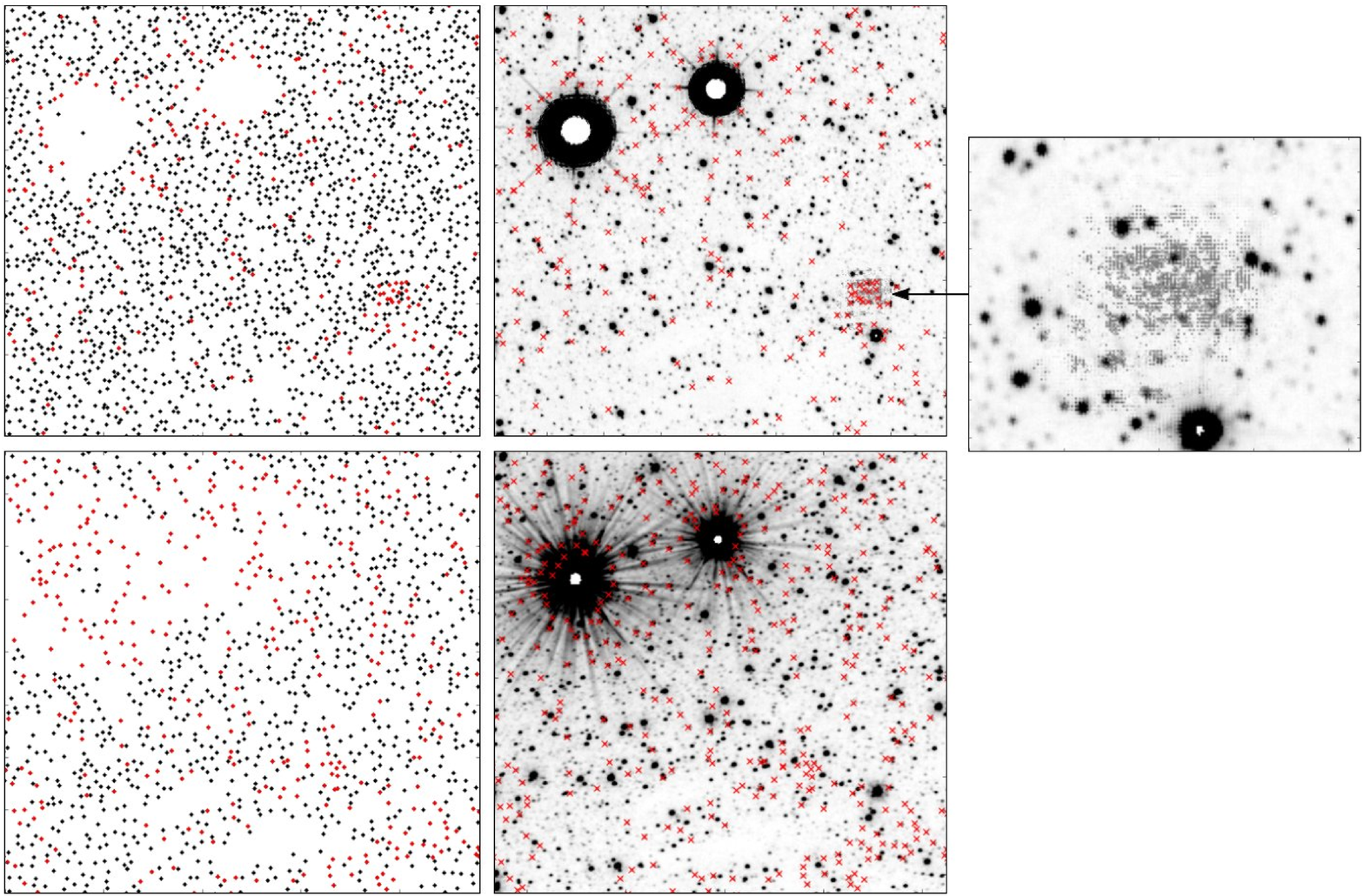}
\caption{Artefacts caused by the $K=3\fm2$ star 2MASS J17571476-2408354 and the $K=4\fm1$ star 2MASS J17571117-2408210. On the upper row are the images and plots from UKIDSS and on the lower row from VVV. In the UKIDSS image is a persistence image at ($l=5.420\degr,b=0.232\degr$). Image orientation is North up and East left. Image size is 2.5\arcmin\ by 2.5\arcmin. Plot markers (dots and crosses) are as in Fig. \ref{brightStar}.}
  \label{persist}%
\end{figure*}

In Fig. \ref{persist} the two bright stars produce a hole into the UKIDSS catalogue and a cluster of non-stellar sources into the VVV catalogue. In Fig. \ref{persist} is also presented the persistence image artefact of UKIDSS. Due to differencies in the telescope and camera optics and observation procedures in the surveys artefacts like the persistence image are not expected to happen at the same locations in the two surveys.

\section{Known clusters}\label{appC}

\begin{figure*}
\centering
\includegraphics[width=\textwidth]{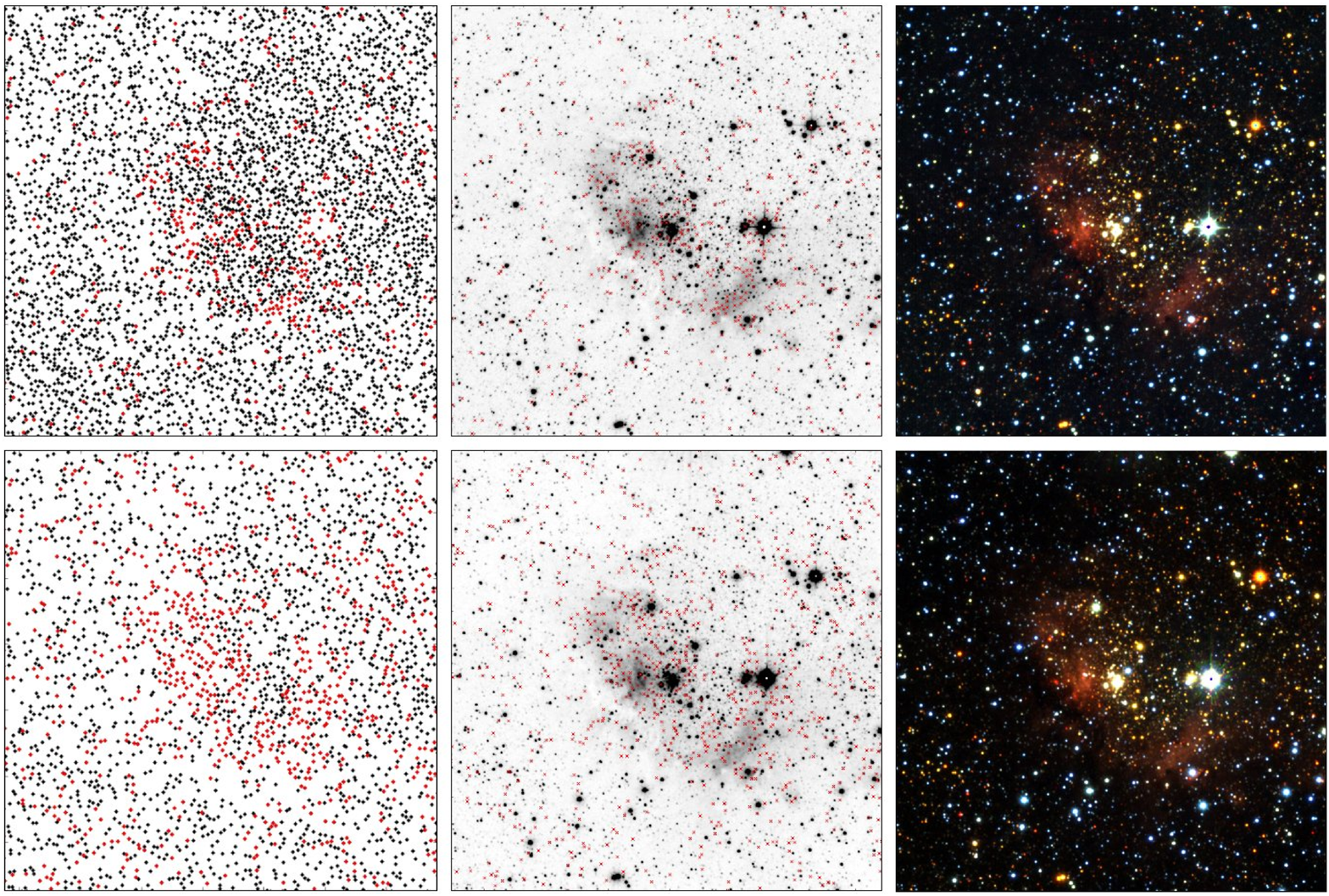}
\caption{Clusters \object{[BDS2003] 113} in the center and \object{[BDS2003] 112} 1.1\arcmin\ southwest from the center. On the upper row are the images and plots from UKIDSS and on the lower row from VVV. In the first column are the catalogue plots, in the middle column the corresponding \textit{K} band image and in the rightmost column false colour images produced from \textit{J}, \textit{H} and \textit{K} band images. Image orientation is North up and East left. Image size is 4\arcmin\ by 4\arcmin. Plot markers (dots and crosses) are as in Fig. \ref{brightStar}.}
  \label{BDS113}%
\end{figure*}

\begin{figure*}
\centering
\includegraphics[width=\textwidth]{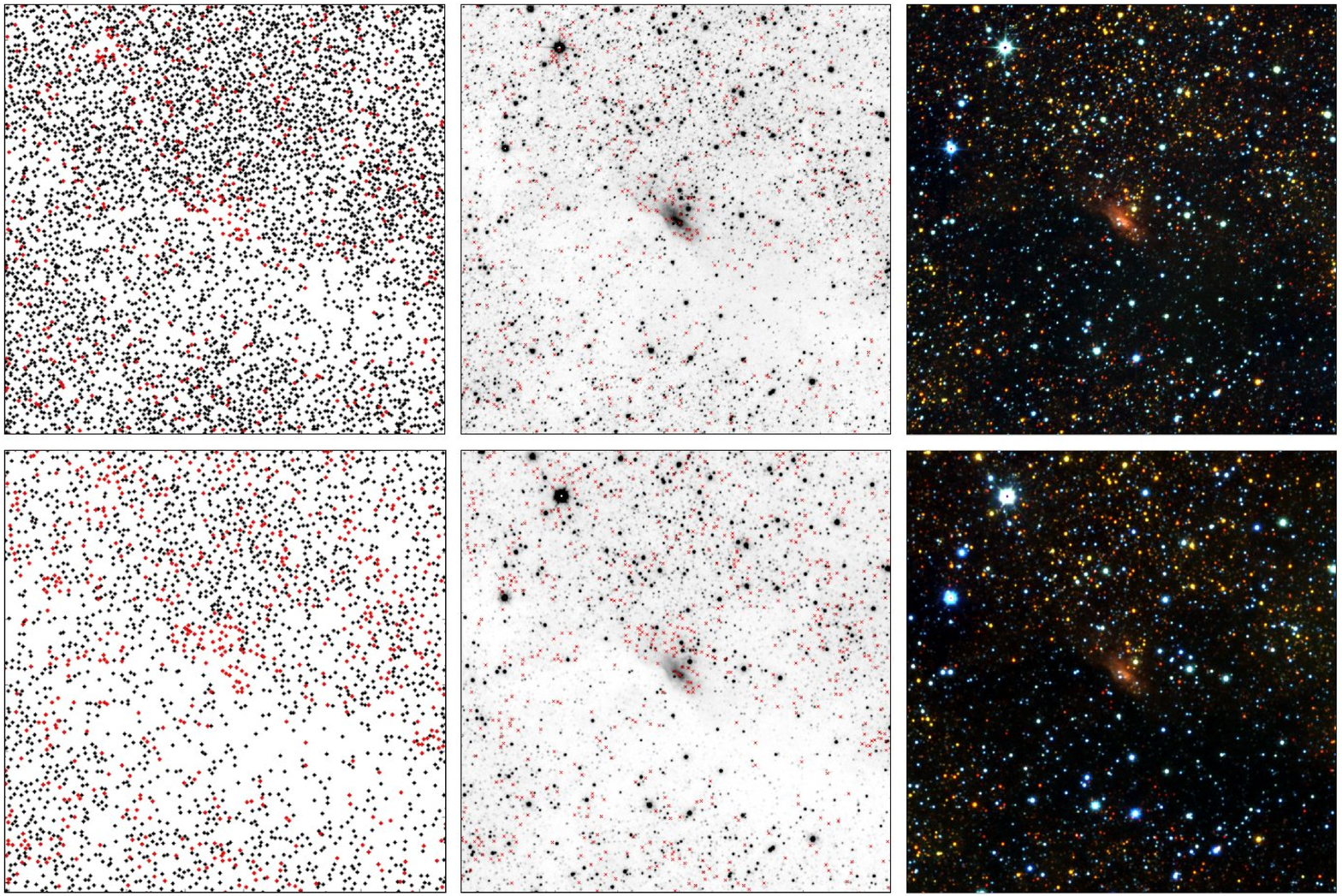}
\caption{As Fig. \ref{BDS113} for cluster \object{[DB2000] 26}.}
  \label{DB26}%
\end{figure*}

\begin{figure*}
\centering
\includegraphics[width=\textwidth]{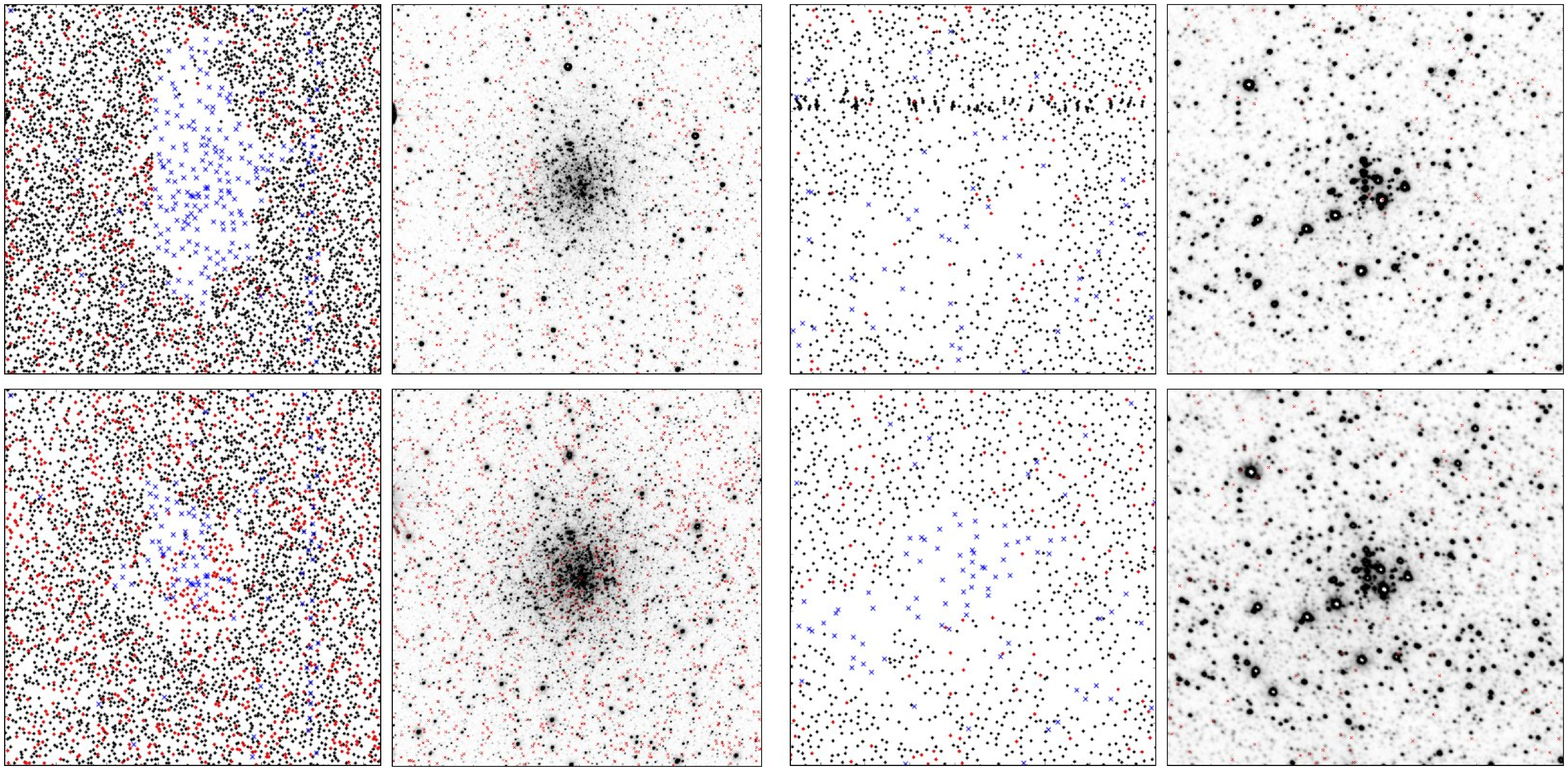}
\caption{Clusters \object{UKS 1751-24.1} in the first two columns and \object{[BDB2003] G000.16-00.06} in the last two columns. On the upper row are the images and plots from UKIDSS and on the lower row from VVV. Image orientation is North up and East left. Image size is 4\arcmin\ by 4\arcmin\ for UKS 1751-24.1 and 2\arcmin\ by 2\arcmin\ for [BDB2003] G000.16-00.06. Blue crosses are sources listed in 2MASS but not in UKIDSS or VVV. Other plot markers (dots and crosses) are as in Fig. \ref{brightStar}.}
  \label{UKS_BDB2}%
\end{figure*}

The longitude range $-2\degr<l<10.4\degr$ is surveyed by both the UKIDSS GPS and the VVV surveys which allows to compare the two data sets (this area is not included in Fig. 1 in Paper I as no cluster candidates were found in this area). A sample of true positive clusters in the fields covered by the both the surveys are shown in Figs. \ref{BDS113}-\ref{UKS_BDB2}. The number of entries in the GPS survey is larger than in the VVV survey but the fraction of sources brighter than $17^\mathrm{m}$ in \textit{K} is larger in the latter. Because of the lower background level (the surplus of faint sources in the GPS survey) the true positive clusters trigger more strongly the search algorithm used in this work.

In Fig. \ref{BDS113} are the clusters \object{[BDS2003] 112} and \object{[BDS2003] 113} and in Fig. \ref{DB26} is the cluster \object{[DB2000] 26}. The UKIDSS catalogue has more entries but the cluster of non-stellar sources is much clearer in the VVV catalogue. The \textit{K} band image and false colour images are very alike for UKIDSS and VVV. For the area around clusters [BDS2003] 112 and 113 98\% of the sources are brighter than $17^\mathrm{m}$ in \textit{K} in the VVV survey catalogue but only 69\% in the UKIDSS catalogue.

In Fig. \ref{UKS_BDB2} are the clusters \object{UKS 1751-24.1} in the first two columns and \object{[BDB2003] G000.16-00.06} in the last two columns. The globular cluster UKS 1751-24.1 produces a hole into the UKIDSS catalogue and a cluster of non-stellar sources into the VVV catalogue. The cluster [BDB2003] G000.16-00.06 produces a hole into the VVV catalogue but leaves no clear traces into the UKIDSS catalogue. Sources listed in 2MASS but not in UKIDSS or VVV are plotted with blue crosses. The 2MASS sources clearly fill the empty spaces in the UKIDSS and VVV survey catalogues. Particularly in the case of UKS 1751-24.1 this bright object produces a fairly large gap into the UKIDSS catalogue.

\end{appendix}

\end{document}